\documentclass[aps,pra,reprint,amsmath,amssymb,superscriptaddress,longbibliography]{revtex4-2} 
\usepackage{color,graphicx,hyperref,float,stackengine,braket,bm,mathtools,calc,mathdots,amsmath}
\usepackage[caption=false]{subfig}
\usepackage{comment}

\begin{document}
\title{Selective cooling and squeezing in a lossy optomechanical closed loop embodying an exceptional surface}
\author{Beyza \surname{S\"{u}tl\"{u}o\u{g}lu Ege}}
\author{Ceyhun \surname{Bulutay}}
\email{bulutay@fen.bilkent.edu.tr}
\affiliation{Department of Physics, Bilkent University, Ankara 06800, Turkey}

\date{\today}

\begin{abstract}
A closed-loop, lossy optomechanical system consisting of one optical and two degenerate mechanical resonators is computationally investigated. This system constitutes an elementary synthetic plaquette derived from the loop phase of the intercoupling coefficients. In examining a specific quantum attribute, we delve into the control of quadrature variances within the resonator selected through the plaquette phase. An amplitude modulation is additionally applied to the cavity-pumping laser to incorporate mechanical squeezing. Our numerical analysis relies on the integration-free computation of steady-state covariances for cooling and the Floquet technique for squeezing. We provide physical insights into how non-Hermiticity plays a crucial role in enhancing cooling and squeezing in proximity to exceptional points. This enhancement is associated with the behavior of complex eigenvalue loci as a function of the intermechanical coupling rate. Additionally, we demonstrate that the parameter space embodies an exceptional surface, ensuring the robustness of exceptional point singularities under experimental parameter variations. However, the pump laser detuning breaks away from the exceptional surface unless it resides on the red-sideband by an amount sufficiently close to the mechanical resonance frequency. Finally, we show that this disparate parametric character entitles frequency-dependent cooling and squeezing, which is of technological importance.
\end{abstract}


\maketitle

\section{Introduction}
The archetypal optomechanical system consists of an optical cavity coupled to a mechanical resonator, commonly mediated by the radiation pressure \cite{marquardtprl2006,aspelmeyer2014cavity,bowen-book}. Its exquisite sensitivity to external forces and displacements makes it attractive for monitoring mechanical motion \cite{kippenberg2007cavity}. Recently, optomechanics thrived in various directions, such as ground-state cooling of mechanical resonators \cite{mancini1998optomechanical,marquardt2007quantum,wilson2007theory,wilson2008cavity,liu2013dynamic,wilson2015measurement, clark2017sideband,rossi2017enhancing}, assorted blockade effects \cite{liao2013photon,ramos2013nonlinear,xie2017phonon,zhang2019enhancing,wang2019distinguishing}, macroscopic entanglement \cite{vitali2007optomechanical,paternostro2007creating}, mechanical squeezing \cite{liao2011parametric,kronwald2013arbitrarily,lu2015squeezed,wollman2015quantum,wang2016steady,wang2016steadydouble}, optomechanically induced transparency \cite{agarwal2010electromagnetically,weis2010optomechanically,safavi2011electromagnetically,chang2011multistability} and high precision measurements \cite{RevModPhys.82.1155}.

A considerable body of these efforts has been devoted to investigating the cooling of mechanical resonators. Various schemes exist for this goal, such as backaction cooling \cite{schliesser2006radiation,massel2012multimode}, and feedback cooling \cite{mancini1998optomechanical,wilson2015measurement}. Another effective choice utilizes the so-called resolved-sideband regime, which is reached when the cavity decay rate is much lower than the frequency of the mechanical oscillator \cite{marquardt2007quantum,wilson2007theory,genes2008ground}. Accordingly, performed experiments revealed a sizable amount of cooling of the mechanical vibrations in this regime \cite{gigan2006self,arcizet2006high,schliesser2006radiation,poggio2007feedback,thompson2008strong,schliesser2008resolved,rocheleau2010preparation}. 
Going beyond a \textit{single} resonator, simultaneous cooling of two mechanical resonators coupled to an optical cavity in the resolved-sideband regime was theoretically suggested \cite{lai2018simultaneous}. It has importance from the standpoint of quantum coherence and also as an experimental realization of cooling of hybridized modes in the mechanical degeneracy regime \cite{ockeloen2019sideband}. However, if two mechanical resonators are coupled to an optical cavity, they form a so-called $\Lambda$ scheme, which generates dark and bright degenerate modes \cite{scully-book}. Though useful for storing quantum information, the dark mode creates an obstacle for multiple cooling \cite{massel2012multimode}. By means of a phase-dependent phonon-exchange loop-coupling, the dark mode effect is alleviated for specific phase values, and multiple mechanical resonators can be cooled to their ground-state \cite{lai2020nonreciprocal}. Alternatively, the dark mode effect can be eliminated for multimode optomechanical cooling by introducing an auxiliary cavity or mechanical mode to the system as a substitute cooling channel \cite{huang2022multimode, lai2022efficient, xu2022millionfold}. 

A closely related endeavor that has gained serious attention is optomechanical \textit{squeezing}. Cavity optomechanics offers a unique platform for generating and manipulating squeezed states of light, which find applications in various fields, including quantum information processing, precision measurements, and gravitational wave detection \cite{braginsky1980quantum,caves1980measurement,meystre2013short}. In connection to the last item, a pressing requirement is the frequency-dependent squeezing tailored to the underlying noise spectrum to facilitate detecting gravitational waves at a certain range of frequencies \cite{chelkowski2005experimental, mcculler2020frequency, junker2022frequency}. Harnessing parametric coupling between optical and mechanical modes \cite{Metelmann-lecture}, optical \cite{clerk2008back,hertzberg2010back} and mechanical \cite{woolley2008nanomechanical,jahne2009cavity,nunnenkamp2010cooling} squeezed states are obtained. Many theoretical proposals are advocated for generating mechanical squeezing \cite{asjad2014robust,lu2015steady,benito2016degenerate}, eventually leading to their experimental realization \cite{tan2013dissipation,kronwald2013arbitrarily,wang2013reservoir,woolley2014two,pirkkalainen2015squeezing,lei2016quantum,ockeloen2018stabilized,barzanjeh2019stationary}. 
A versatile tool in this direction is periodic modulation which is also used for intentions, such as enhancing quantum effects  \cite{farace2012enhancing}, and generation of entanglement \cite{chen2014enhancement,wang2016macroscopic}, exploiting thermo-optic nonlinearity \cite{pelkaprl2022} and achieving phonon blockade and nonclassical states \cite{yin2017nonlinear}, apart from generation of mechanical squeezing \cite{mari2009gently,liao2011parametric,hanpra2019,bai2020strong}.
One blueprint for the latter is two-tone driving with both red- and blue-detuned lasers \cite{kronwald2013arbitrarily}, where the mechanical squeezing can be detected by directly measuring the cavity output spectrum \cite{huangpra2021}.

A recently flourishing, yet seemingly disparate topic is artificial gauge potentials \cite{dalibardrmp11,umucalilarpra11}, bestowing features such as nonreciprocal photon transport \cite{fang2017generalized,huangprl2018,lepinay2020}, on-chip optical nonreciprocity, and optical isolation \cite{hafezi2012optomechanically,fang2012photonic}. Optical nonreciprocity refers to the phenomenon in which the transmission of light is different in one direction compared to the opposite, which yields various applications such as isolators \cite{jalas2013and}, circulators \cite{fleury2014sound}, and directional amplifiers \cite{de2019realization}. 
In optomechanics, nonreciprocity offers a promising avenue for manipulating thermal fluctuations, such as for cooling phononic resonators \cite{xu2019nonreciprocal}, thermal-noise cancellation \cite{tang2023}, mode routing and thermal management \cite{shenprl2023}. Such nonreciprocity can be created by breaking the time-reversal symmetry in the system, which is one of the prime uses of synthetic gauge fields \cite{xu2020nonreciprocity}. 
A frequency difference between mechanical modes or cavity field modulation is a reliable method for creating synthetic fields \cite{schmidt2015optomechanical,mathew2020synthetic}. They enable tunable cooling in the optomechanical system by introducing a phase-dependent coupling constant \cite{lai2020nonreciprocal,jiang2021energy}. Most recently, strong mechanical squeezing has been proposed by breaking the dark-mode effect with the synthetic-gauge-field \cite{huang2023controllable}.

Captivating all these fields, a revolutionary wave of non-Hermitian physics is currently underway, offering unique advantages empowered by the so-called exceptional point (EP) singularities \cite{benderbook,el2018non,ozdemir2019parity}. Naturally, non-Hermitian phenomena have become an appealing resource, so far largely capitalized in photonics, 
with unprecedented consequences such as, loss-induced revival of lasing \cite{pengscience2014}, phonon lasing \cite{zhang2018phonon}, boosting optomechanical interactions with high-order EPs \cite{jing2017high}, enhanced quantum sensing \cite{lau2018fundamental}. 
As a few other outstanding yet highly relevant studies for our work, mechanical cooling is proposed in a non-Hermitian system \cite{liao2023exceptional}, also with a synthetic gauge field \cite{jiang2021energy}, where the average phonon occupation number is minimized at the EP, and chiral, non-Hermitian tunable mechanical squeezing in nano-optomechanical networks are experimentally realized \cite{del2022non}. Another notable development in this front is the concept of an \textit{exceptional surface} on which the EPs move \cite{zhongprl19}. Its presence assures insensitivity to variations over the parameters that define the exceptional surface. As such, the existence of an EP does not require unrealistic fine-tuning of parameters. Soon after the exceptional surfaces were proposed \cite{zhou19,zhongol19}, they were experimentally demonstrated for two-mode systems of magnon polariton \cite{zhang19} and optical waveguides \cite{soleymani22}.

In line with this progress, the aim of this work is to present a unified treatment of both cooling and squeezing of the mechanical quadratures controllable through a closed-loop phase of a photonic cavity coupled to two intercoupled lossy mechanical resonators.
Compared to our recent work, where we studied loop phase control of optomechanically-induced transparency in a gain-loss non-Hermitian setting, here we tackle the control of \textit{quantum} phenomena of quadrature variances and squeezing, in contrast to probe transmission, which is essentially governed by \textit{classical} mean fields \cite{beyza21}.
Our numerical framework constitutes the Floquet formalism \cite{shirleypr1965,malzpra2016,pietikainen2020combining} for steady-state or modulated evolution. First, mechanical cooling is explored based on an experimentally attainable parameter set. Moreover, mechanical squeezing can be accomplished by imposing an upper-sideband amplitude modulation of the cavity pumping laser while the loop-phase selects the resonator with maximum squeezing. We demonstrate that non-Hermiticity plays an intricate role, as determined by the proximity to two EPs derived from the upper- and lower-frequency polariton-phonon modes. Next, we show that the parameter variations of this three-mode optomechanical system guide the EPs over an exceptional hypersurface, making them robust against fabrication errors and fluctuations in the experimental environment. The exception is the detuning, as it needs to be on the red-sideband by an amount close to the mechanical resonance frequency; otherwise, with further deviation, the system breaks away from the exceptional surface. Favorably, this behavior bestows a frequency-dependent squeezing as a function of laser detuning.    

The paper is organized as follows. In Sec.~II, we introduce our model and present its theoretical analysis. In Sec.~III, we discuss possible experimental realization schemes and our choice of parameters. This is followed by our results in Sec.~IV, divided into three subsections: mechanical cooling, exceptional surfaces, and squeezing. Our main conclusions are drawn in Sec.~V. Appendix~A  contains the steady-state values of the cavity and mechanical modes; Appendix~B derives the covariance matrix expressions for the Floquet formalism; Appendix~C derives integration-free computation of steady-state variances.

\section{Theory}
\subsection{Optomechanical Network Hamiltonian}
\begin{figure}[H]
  \centering
  \includegraphics[width=0.5\textwidth]{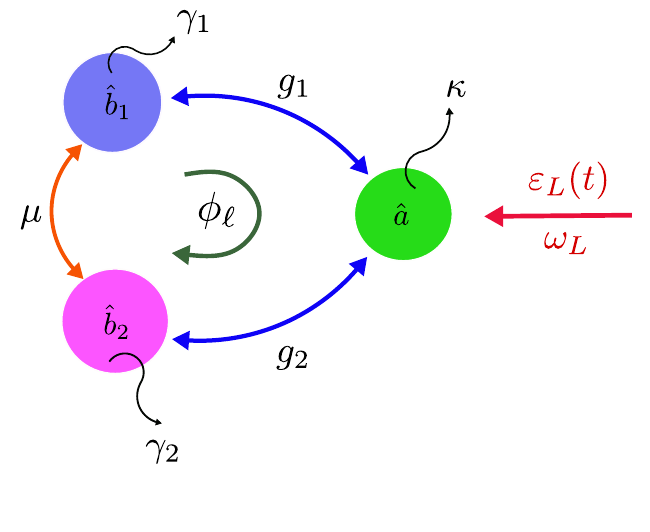}
\caption{Closed-loop interaction optomechanical system composed of a photonic cavity with the relevant resonance at $\omega_{cav}$, and two mechanical resonators with identical frequencies, $\omega_m$. Loss rates are indicated with wavy arrows. The cavity is pumped with a laser (in the case of squeezing also modulated) having a carrier frequency $\omega_L$ and amplitude $\varepsilon_L(t)$.}
\label{fig1}
\end{figure}

We consider the simplest closed optomechanical network consisting of a photonic cavity coupled to two mechanical resonators via coupling rates $g_1$ and $g_2$ as shown in Fig~(\ref{fig1}). The mechanical resonators are bilinearly intercoupled with a rate $\mu$, and both have equal resonance angular frequency $\omega_m$, and damping rates $\gamma_1$ and $\gamma_2$. The photonic cavity is driven, in the most general case, with an amplitude-modulated laser with carrier frequency $\omega_L$ and amplitude $\varepsilon_L(t)$; when applied, the modulation frequency is $\Omega=2\pi/\tau$ so that 
$\varepsilon_L(t+\tau) = \varepsilon_L(t) = {\sum_{n=-\infty}^{\infty}\varepsilon_n}e^{-in\Omega t}$, with $d_n=\varepsilon_n/\varepsilon_0$ being the modulation amplitude depth associated with the sideband $n$ where amplitude  $\varepsilon_0$ is defined in terms of the laser power, $\varepsilon_0 = \sqrt{\frac{2P}{\hbar \omega_L}}$.

The Hamiltonian ($\hbar=1$) in the rotating frame of the pump laser at frequency $\omega_L$ is
\begin{eqnarray}
\hat{H} & = &  \Delta \hat{a}^\dagger \hat{a} + \omega_m (\hat{b}_1^\dagger \hat{b}_1 + \hat{b}_2^\dagger \hat{b}_2)- (\mu\hat{b}_1^\dagger \hat{b}_2 + \mu^*\hat{b}_2^\dagger \hat{b}_1)\nonumber \\
& & -\hat{a}^\dagger \hat{a}(g_1\hat{b}_1^\dagger + g_1^*\hat{b}_1) -  \hat{a}^\dagger \hat{a}(g_2 \hat{b}_2^\dagger + g_2^* \hat{b}_2) \nonumber\\
& &+ i \sqrt{\eta \kappa} \left[\varepsilon_L(t) \hat{a}^\dagger-\varepsilon^*_L(t) \hat{a}\right],
\label{Hamiltonian}
\end{eqnarray}
where $\Delta= \omega_{cav}-\omega_L$ is the frequency detuning between the cavity and input laser, $\hat{a}$ ($\hat{a}^\dagger$), and $\hat{b}_1$ ($\hat{b}_1^\dagger$) and $\hat{b}_2$ ($\hat{b}_2^\dagger$) are the annihilation (creation) operators of the optical, first and second mechanical resonator modes, respectively, $\kappa$ is the cavity decay rate, and $\eta$ is the cavity coupling parameter (escape efficiency). The non-Hermiticity of the system follows from the lossy nature of all the resonators.

\subsection{Closed-loop phase}
\begin{figure}[H]
  \centering
  \includegraphics[width=0.5\textwidth]{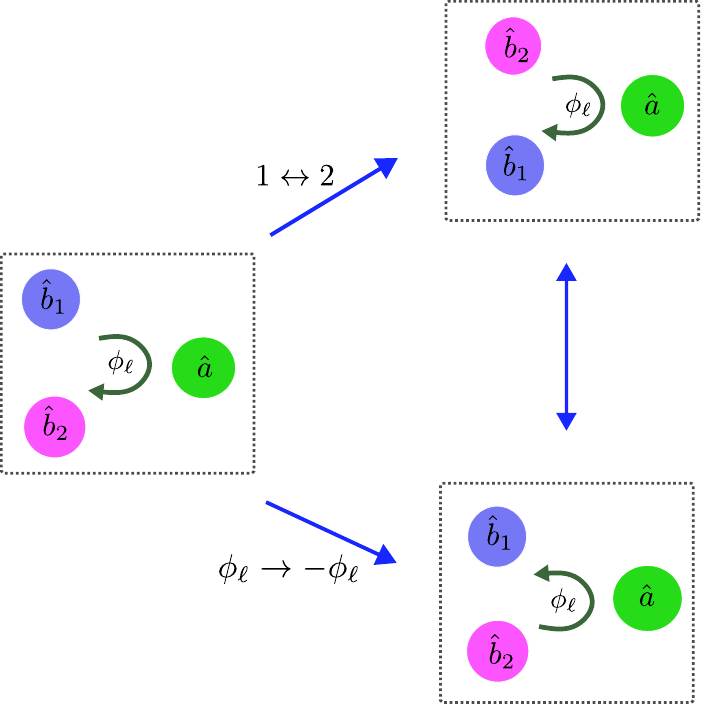}
\caption{In the closed-loop coupling scheme, swapping the two identical mechanical resonators $1\longleftrightarrow 2$, is equivalent to $\phi_\ell\rightarrow 2\pi-\phi_\ell$.}
\label{fig-symm}
\end{figure}

The three complex coupling coefficients, $g_1=|g_1|\, e^{i\phi_1}$, $g_2=|g_2|\, e^{i\phi_2}$, and $\mu=|\mu|\, e^{i\phi_\mu}$, in Eq.~(\ref{Hamiltonian}) comprise a closed-loop phase, $\phi_\ell\coloneqq -\phi_1+\phi_2+\phi_\mu$ \cite{beyza21}. However, as we discuss in Sec.~III, retaining $g_1, g_2\in \mathbb{R}$ while imparting $\phi_\mu\rightarrow\phi_\ell$ is the viable implementation from a practical standpoint. Notably, $\phi_\ell$, just like a Peierls phase, threads synthetic magnetic flux through this triangle plaquette \cite{chen2021synthetic}. $\phi_\ell$ is a gauge-invariant quantity having observable consequences as it controls which mechanical resonator is to be predominantly cooled or squeezed. In the case of \textit{identical} mechanical resonators, a crucial consequence of this choice is its invariance under simultaneous swapping of the two mechanical resonators, $1\longleftrightarrow 2$ along with $\phi_\ell\rightarrow -\phi_\ell=2\pi-\phi_\ell$ (see Fig.~\ref{fig-symm}). In other words, one can toggle between the selected mechanical resonators via a phase reversal. We expound on this symmetry in the Results sections.

\subsection{Quantum Langevin equations}
To account for input noise and losses, we switch to quantum Langevin equations \cite{gardiner-zoller}, which describe the dynamics of the cavity and mechanical modes as 
\begin{eqnarray*}
\frac{d\hat{a} }{dt}&=& -i\Delta \hat{a} + i \hat{a}(g_1 \hat{b}_1^\dagger + g_1^* \hat{b}_1) + i\hat{a} (g_2 \hat{b}_2^\dagger + g_2^* \hat{b}_2) \nonumber \\
& & + \sqrt{\eta\kappa} \varepsilon_L(t)-\frac{\kappa}{2}\hat{a} +\sqrt{\kappa}\hat{a}_{in}(t),\\
\frac{d\hat{b}_1}{dt} &=& -i\omega_m \hat{b}_1 +i\mu\hat{b}_2 +i g_1 \hat{a}^\dagger \hat{a} -\frac{\gamma_1}{2} \hat{b}_1 +\sqrt{\gamma}_1 \hat{b}_{1,in}(t),\\
\frac{d\hat{b}_2}{dt} &=& -i\omega_m \hat{b}_2 +i\mu^*\hat{b}_1 +i g_2 \hat{a}^\dagger \hat{a} -\frac{\gamma_2}{2} \hat{b}_2 +\sqrt{\gamma}_2 \hat{b}_{2,in}(t),\\
\nonumber
\end{eqnarray*}
where $\hat{a}_{in}(t)$, $\hat{b}_{1,in}(t)$ and $\hat{b}_{2,in}(t)$ are zero-mean cavity and mechanical input noise operators, respectively. They satisfy the following correlation functions (displaying only the non-zero ones) under the Markovian-reservoir assumption \cite{RevModPhys.82.1155},
\begin{eqnarray}
\langle \hat{a}_{in}^\dagger(t) \hat{a}_{in}(t') \rangle &=& n_a \delta(t-t'),\\
\langle \hat{a}_{in}(t) \hat{a}_{in}^\dagger(t') \rangle &=& (n_a +1) \delta(t-t'),\\
\langle \hat{b}_{j,in}^\dagger(t) \hat{b}_{j,in}(t') \rangle &=& n_m \delta(t-t'),\\
\langle \hat{b}_{j,in}(t) \hat{b}_{j,in}^\dagger(t') \rangle &=& (n_m+1) \delta(t-t'),
 \end{eqnarray}
where $j=1,2$, $n_a$, and $n_m$ are the mean occupancy of the cavity and mechanical baths, respectively. For cooling and squeezing purposes, our primary focus is on quantum fluctuations. Thus, we linearize the system by writing cavity mode $\hat{a}$ and mechanical modes $\hat{b}_1$, $\hat{b}_2$ as a sum of the classical mean value and quantum fluctuations operators $\hat{\aleph}(t)\rightarrow \langle\hat{\aleph}(t)\rangle + \delta\hat{\aleph}(t)$. The remaining details of the steady-state values are provided in Appendix~A.

The quantum fluctuations around the classical mean values, represented by $\delta\hat{\aleph}(t)$, have the following equations of motion
\begin{eqnarray*}
\label{qfcav}
\frac{d\delta\hat{a}}{dt}&=& i(-\Delta_a+i\kappa/2)\delta\hat{a}+\sqrt{\kappa}\hat{a}_{in}(t)+i\langle \hat{a} \rangle g_1^* \delta \hat{b}_1 \\
& & +i\langle \hat{a}\rangle g_1 \delta \hat{b}_1^\dagger +i \langle \hat{a} \rangle g_2^* \delta \hat{b}_2 +i\langle \hat{a} \rangle g_2 \delta \hat{ b}_2^\dagger,\\
\label{qfMR1}
\frac{d\delta \hat{b}_1}{dt} &=& i(-\omega_m +i\gamma_1/2)\delta \hat{b}_1 +i\mu\delta \hat{b}_2 + ig_1\langle \hat{a} \rangle^* \delta \hat{a} \\ 
& & + ig_1\langle \hat{a} \rangle \delta \hat{a}^\dagger + \sqrt{\gamma_1} \hat{b}_{1,in}(t),\\
\label{qfMR2}
\frac{d\delta \hat{b}_2}{dt} &=& i(-\omega_m +i\gamma_2/2)\delta \hat{b}_2 +i\mu^*\delta \hat{b}_1 + ig_2\langle \hat{a} \rangle^* \delta \hat{a} \\
& & + ig_2\langle \hat{a} \rangle \delta \hat{a}^\dagger + \sqrt{\gamma_2} \hat{b}_{2,in}(t).
\label{quantum fluctuation}
\end{eqnarray*}
Next, we switch to experimentally accessible quadrature operators of position and momentum, which are expressed in terms of the fluctuation operators as 
\begin{eqnarray*}
\delta \hat{X}_{\aleph=a,b_1,b_2} &=&  \frac{\delta \hat{\aleph} +\delta\hat{\aleph}^\dagger }{\sqrt{2}}, \\
\delta \hat{Y}_{\aleph=a,b_1,b_2} &=&  \frac{\delta \hat{\aleph} -\delta\hat{\aleph}^\dagger }{i \sqrt{2}},
\end{eqnarray*}
and the corresponding quadrature noise operators are
\begin{eqnarray*}
\hat{X}^{in}_{\aleph=a,b_1,b_2} &=&  \frac{\hat{\aleph}_{in} +\hat{\aleph}_{in} ^\dagger }{\sqrt{2}},\\
\hat{Y}^{in}_{\aleph=a,b_1,b_2} &=&  \frac{\hat{\aleph}_{in} -\hat{\aleph}_{in} ^\dagger }{i \sqrt{2}} \, .
\end{eqnarray*}

Finally, the position-momentum quadrature fluctuations of the cavity and mechanical modes can be cast in the form \cite{bai2020strong} 
\begin{equation}
\label{quadraturefluc eqn}
\mathbf{\dot{\hat{R}}}(t)=\mathbf{M}(t) \mathbf{\hat{R}}(t)+ \mathbf{\hat{N}}(t),  
\end{equation}
where $\mathbf{\hat{R}}(t)=[\delta \hat{X}_{a},\delta\hat{Y}_{a},\delta\hat{X}_{b_1},\delta\hat{Y}_{b_1},\delta\hat{X}_{b_2},\delta\hat{Y}_{b_2}]^T$, with the superscript $T$ indicating vector or matrix transpose and the dot, $\mathbf{\dot{\hat{R}}}$ represents time derivative.
$\mathbf{M}(t)$ is a 6$\times$6 time-dependent matrix with its elements (not included here) readily extracted from the equations of motion of $\delta\hat{\aleph}(t)$. $\mathbf{\hat{N}}(t)$ is the noise operator vector and defined as $$\mathbf{\hat{N}}(t)=\left[\sqrt{\kappa}\hat{X}^{in}_{a},\sqrt{\kappa}\hat{Y}^{in}_{a}, \sqrt{\gamma_1}\hat{X}^{in}_{b_1},\sqrt{\gamma_1}\hat{Y}^{in}_{b_1},\sqrt{\gamma_2}\hat{X}^{in}_{b_2},\sqrt{\gamma_2}\hat{Y}^{in}_{b_2}\right]^T\, .$$
We will solve Eq.~(\ref{quadraturefluc eqn}), which is a first-order inhomogeneous differential equation. 
\\
\subsection{Quadrature variances}
The formal solution of Eq.~(\ref{quadraturefluc eqn}) is
\begin{equation}
\mathbf{\hat{R}}(t) = \mathbf{G}(t)\mathbf{\hat{R}}(0) + \mathbf{G}(t)  \int_{0}^{t} \mathbf{G}^{-1}(\tau) \mathbf{\hat{N}}(\tau) d\tau,  
\end{equation}
where $\mathbf{G}(t)$ satisfies $\mathbf{\dot{G}}(t)=\mathbf{M}(t)\mathbf{G}(t)$ subject to the initial condition $\mathbf{G}(0)= \mathbf{I}$, with $\mathbf{I}$ being the identity matrix \cite{bai2020strong}. 
In order to investigate the mechanical squeezing and cooling, we need quadrature fluctuations of mechanical resonators. Hence, we introduce the covariance matrix $\mathbf{V}(t)$ to analyze the dynamics of this optomechanical system where its entries are defined as
\begin{equation}
\label{CM}
\mathbf{V}_{ij}(t) = \langle \mathbf{\hat{R}}_{i}(t) \mathbf{\hat{R}}_{j}(t) \rangle ,    
\end{equation}
for $i,j=1,2,\ldots,6$. From the definition of the covariance matrix and Eq.~(\ref{CM}), we obtain 
\begin{equation}
\mathbf{V}(t) = \mathbf{G}(t)\mathbf{V}(0)\mathbf{G}^T(t) + \mathbf{G}(t)\mathbf{S}(t)\mathbf{G}^T(t),
\end{equation}
where 
\begin{equation}
\mathbf{S}(t) = \int_{0}^{t}  \int_{0}^{t} \mathbf{G}^{-1}(\tau) \mathbf{K}(\tau,\tau^{'})\left[\mathbf{G}^{-1}(\tau)\right]^T d\tau d\tau^{'},
\end{equation}
in which $ \mathbf{K}(\tau,\tau^{'})$ is two-time noise correlation function whose elements are $ \mathbf{K}(\tau,\tau^{'})= \langle \mathbf{\hat{N}}(\tau) \mathbf{\hat{N}}(\tau^{'})\rangle= \mathbf{C} \delta(\tau-\tau^{'})$, where 
\begin{widetext}
\begin{equation}
\mathbf{C}=
\begin{pmatrix}
\frac{\kappa}{2}(2n_a+1) & \frac{-\kappa}{2i}&0&0&0&0 \\
\frac{\kappa}{2i} & \frac{\kappa}{2}(2n_a+1)  &0&0&0&0 \\
0&0& \frac{\gamma_1}{2}(2n_m+1)&\frac{-\gamma_1}{2i}&0&0\\
0&0&  \frac{\gamma_1}{2i}&\frac{\gamma_1}{2}(2n_m+1)&0&0\\
0&0&0&0&\frac{\gamma_2}{2}(2n_m+1)&\frac{-\gamma_2}{2i} \\
0&0&0&0&\frac{\gamma_2}{2i}&\frac{\gamma_2}{2}(2n_m+1) \\
\end{pmatrix}.
\end{equation}
\end{widetext}
Entries $\mathbf{V}_{33}$, $\mathbf{V}_{44}$, $\mathbf{V}_{55}$ and $\mathbf{V}_{66}$ give the position and momentum variances of first and second mechanical resonators, respectively. Squeezing beyond the vacuum state occurs when $-10\log_{10}[V_{ii}/0.5]>0$~dB, since for the vacuum state has $  \delta X^2  =  \delta Y^2 =0.5$~. 
\subsection{Floquet Analysis}
The formal solution gives the variance as a function of time. An alternative is the Floquet method, which yields variance when $t \rightarrow \infty$, i.e., in the steady state regime.
Due to the modulation with frequency $\Omega$, in the absence of noise the dynamical variables, $\mathbf{\hat{R}}(t)$, are also periodic with $\Omega$ in the steady-state i.e., $\mathbf{\hat{R}}(t) = \mathbf{\hat{R}}(t+2\pi/\Omega)$. This allows us to expand these dynamical variables into the Fourier series 
\begin{equation}
\mathbf{\hat{R}}(t) =  \sum_{n=-\infty}^{\infty} \mathbf{\hat{R}}^{(n)}(t)  e^{-in\Omega t},
\end{equation}
whereas Fourier coefficients are actually time-dependent due to noise. Introducing the Fourier transformation for these Fourier series coefficients, we obtain
\begin{equation}
\mathbf{\hat{R}}^{(n)}(t) = \frac{1}{2\pi} \int_{-\infty}^{\infty} \mathbf{\hat{R}}^{(n)}(\omega) e^{-i\omega t} d\omega.
\end{equation}
We can also expand the so-called drift matrix $\mathbf{M}(t)$  into its harmonics as
\begin{equation}
\mathbf{M}(t)= \mathbf{M}^{(1)} e^{-i\Omega t} + \mathbf{M}^{(0)} + \mathbf{M}^{(-1)} e^{i\Omega t}.
\end{equation}
Eq.~(\ref{quadraturefluc eqn}) becomes 
\begin{widetext}
\begin{eqnarray}
\label{eqnfourier}
\nonumber
\sum_{n=-\infty}^{\infty} \frac{d}{dt} \bigg(\frac{1}{2\pi}  \int_{-\infty}^{\infty}  \mathbf{\hat{R}}^{(n)}(\omega) e^{-i\omega t} e^{-in\Omega t} d\omega \bigg) &=&   \sum_{n=-\infty}^{\infty} \sum_{j=-1,0,1} \mathbf{M}^{(j)} e^{-ij\Omega t} \frac{1}{2\pi} \int_{-\infty}^{\infty}  \bigg( \mathbf{\hat{R}}^{(n)}(\omega) \times  \\
\nonumber
& &e^{-in\Omega t}   + \mathbf{\hat{N}}[\omega] \delta_{n,0} \bigg) e^{-i\omega t} d\omega.
\end{eqnarray}
Using $ \int_{-\infty}^{\infty} \mathbf{\hat{R}}^{(n)}(\omega) e^{-in\Omega t} e^{-i\omega t} d\omega = \int_{-\infty}^{\infty} \mathbf{\hat{R}}^{(n)}(\omega-n\Omega) e^{-i\omega t}$ on both sides of the previous equation yields
\begin{eqnarray}
\nonumber
\frac{1}{2\pi} \int_{-\infty}^{\infty} \bigg( (-i\omega) \mathbf{\hat{R}}^{(n)}(\omega-n\Omega)\bigg) e^{-i\omega t}d\omega & = & \frac{1}{2\pi} \int_{-\infty}^{\infty} e^{-i\omega t} d\omega \bigg( \mathbf{M}^{(1)} \mathbf{\hat{R}}^{(n)}(\omega-(n+1)\Omega)   \\
\nonumber
& &+ \mathbf{M}^{(0)} \mathbf{\hat{R}}^{(n)}(\omega-n\Omega)  \\
\nonumber
& &+ \mathbf{M}^{(-1)} \mathbf{\hat{R}}^{(n)}(\omega-(n-1)\Omega) \bigg).
\nonumber
\end{eqnarray}
For this equality to hold, terms inside the parenthesis on the left and right-hand sides must be equal, where $\mathbf{\hat{R}}^{n}(\omega-\Omega)=\mathbf{\hat{R}}^{n-1}(\omega)$. In this case, we have the following equality for Eq.~(\ref{quadraturefluc eqn}), 
\begin{equation}
\label{general Floquet of CM}
\mathbf{M}^{(1)} \mathbf{\hat{R}}^{(n-1)}(\omega) + \bigg(i(\omega+n\Omega)\mathbf{I}+ \mathbf{M}^{(0)}\bigg) \mathbf{\hat{R}}^{(n)}(\omega) + \mathbf{M}^{(-1)} \mathbf{\hat{R}}^{(n+1)}(\omega) = -\delta_{n,0} \mathbf{\hat{N}}(\omega).
\end{equation}
Here, the advantage is the removal of time dependence in the $\mathbf{M}$ matrix. Now, we have a time-independent but infinitely coupled set of algebraic equations among different harmonic contributions. We write Eq.~(\ref{general Floquet of CM}) as
\begin{equation}
\label{inf matrix}
\hspace*{-1cm}
\begin{pmatrix}
\ddots &\vdots &\vdots &\vdots &\vdots & \vdots &\iddots \\
\dots &\mathbf{M}^{(1)} & i(\omega-\Omega)\mathbf{I} + \mathbf{M}^{(0)} &  \mathbf{M}^{(-1)} & \dots &\dots &\dots\\
\dots & \dots& \mathbf{M}^{(1)} & i\omega \mathbf{I} + \mathbf{M}^{(0)} & \mathbf{M}^{(-1)} & \dots&\dots \\
\dots & \dots & \dots& \mathbf{M}^{(1)} & i(\omega+\Omega)\mathbf{I} + \mathbf{M}^{(0)} & \mathbf{M}^{(-1)} & \dots \\
\iddots &\vdots &\vdots &\vdots &\vdots & \vdots &\ddots \\
\end{pmatrix}
\begin{pmatrix}
\vdots \\
\mathbf{\hat{R}}^{(-2)} \\
\mathbf{\hat{R}}^{(-1)} \\
\mathbf{\hat{R}}^{(0)} \\
\mathbf{\hat{R}}^{(1)} \\
\mathbf{\hat{R}}^{(2)} \\
\vdots \\
\end{pmatrix}
=
\begin{pmatrix}
\vdots \\
\mathbf{0}\\
\mathbf{0} \\
-\mathbf{\hat{N}}(\omega) \\
\mathbf{0} \\
\mathbf{0} \\
\vdots \\
\end{pmatrix} 
\end{equation}
\end{widetext}
Eq.~(\ref{inf matrix}) can be written as $\mathbf{P}(\omega) \mathbf{\hat{R}}(\omega)=\mathbf{\hat{n}}(\omega)$. 
Inevitably, we need to limit Fourier components to a maximum size, $\pm N$, which we take as $N=2$ after checking its convergence. Forthwith, we can solve $\mathbf{\hat{R}}(\omega)$ by matrix inversion as $\mathbf{\hat{R}}(\omega)= \mathbf{T}(\omega)\mathbf{\hat{n}}(\omega)$ where $\mathbf{T}(\omega)=\mathbf{P}^{-1}(\omega)$. Importantly, the entries of $\mathbf{P}$ in Eq.~(\ref{inf matrix}) are also matrices, and the overall dimension is $6(2N+1) \times 6(2N+1) $. Dimension of $\mathbf{\hat{R}}(\omega)$ is $6 \times (2N+1)$. In Appendix~B, we extend this formalism to the covariance matrix $\mathbf{V}(t)$. To limit the technical level of the main text, the remaining expressions of this formalism
are deferred to Appendix~B. Moreover, in Appendix~C, we provide an integration-free calculation of steady-state variances in the absence of modulation.
\section{Possible Experimental Realization and the Parameter Set}
In order to substantiate the applicability of our work, we first argue on its possible experimental realization.
Here, the main challenge is achieving the closed-loop phase, $\phi_\ell$.
This being a gauge invariant quantity, it can be imparted to any one of the couplings or divided among them
while leading to the same physical consequence \cite{beyza21}. We believe that the most viable option is incorporating the plaquette 
phase solely to the intermechanical coupling $\mu$, where it is mediated by an optical link with its phase being locked 
to the optical cavity pumping laser up to an offset that is tunable by electro-optic and thermo-optic means \cite{zhang12,gilsantos17}.

Among numerous existing experimental studies pertaining to our system \cite{safavi2011electromagnetically,massel2012multimode,ockeloen2018stabilized,xu2019nonreciprocal,lepinay2020,del2022non}, we pick our parameter set directly from the seminal work of Safavi-Naeini et al. \cite{safavi2011electromagnetically}. However, this work having a single mechanical resonator lacks the intermechanical coupling constant, $\mu$. For our purposes, we take this liberty to vary the $\mu$ values around $\mu_\mathrm{EP,1,2}$ in order to probe the intriguing physics occurring close to EPs. 
To simplify the parameter space and also to benefit from the intrinsic symmetry illustrated in Fig.~\ref{fig-symm}, we consider \textit{identical} mechanical resonators. In accordance with Ref.~\cite{safavi2011electromagnetically} we choose $g_{1,2}\in\mathbb{R}$ with $g_1/(2\pi)=g_2/(2\pi)=800$~kHz, $\omega_m/(2\pi)=3.75$~GHz, $\kappa/(2\pi)=900$~MHz, $\gamma_1=\gamma_2= 5 \times 10^{-4}\,\omega_m$, optical cavity escape efficiency $\eta=0.5$, laser power $P$ = 0.25~mW with a wavelength of 1550~nm, and detuning $\Delta = \omega_m$ (red-detuned pumping), and modulation angular frequency $\Omega=2\omega_m$. This secures the resolved sideband regime because of $\kappa<\omega_m$. 
Unless stated otherwise, the ambient cavity and mechanical reservoir temperatures are taken as $T = 18.1$~K. 
Having $g_{1,2}\ll \kappa$, ensures that the optomechanical system is in the easily attainable weak optomechanical coupling limit. On the other hand, $\mu \gg \gamma_{1,2}$ indicates the strong intermechanical coupling regime. Especially reaching the $\mu_\mathrm{EP,1,2}$ values is experimentally rather formidable. Nevertheless, with the recent progress, strong intermechanical coupling regime is becoming more accessible \cite{okamoto2013coherent, Mathew_2016,  deng2016strongly, luo2018strong}.

We should note that this full parameter set was not pre-optimized for mechanical cooling or squeezing purposes \cite{safavi2011electromagnetically}. In analyzing any of these parameters' sensitivity, we designate the numerical values above with subscripts such as $\kappa_c ,~P_c$, etc. For this data set, there are two EPs for the intermechanical coupling rate, $\mu$ occurring at $|\mu_\mathrm{EP,1}|\simeq 52.5\,(\gamma_1+\gamma_2)$ and $|\mu_\mathrm{EP,2}|\simeq 80.45\,(\gamma_1+\gamma_2)$, which we utilize in our following analysis. For each $|\mu_\mathrm{EP}|$ there are two distinct $\phi_\ell$ values corresponding to opposite chirality EPs, such that the one with $\phi_\ell<\pi$ is the clockwise and $\phi_\ell>\pi$ is the counterclockwise EP.
Having $\kappa>\gamma_{1,2}>0$, this makes up a \textit{loss-loss} system. As shown by \"{O}zdemir et al., as long as the mode losses are \textit{nonuniform} (as in our case) under a gauge transformation, it can be mapped to the prototypical \textit{gain-loss} non-Hermitian model \cite{ozdemir2019parity}.
\section{Results}
\subsection{Mechanical Cooling}
We begin our analysis with the optomechanical cooling process. Under red-detuned pumping, the laser photons to be admitted into the cavity need to gain the deficient energy from the phonons of the mechanical resonators, specifically through the $\hat{a}^\dagger\hat{b}_{1,2}$ terms of the Hamiltonian [cf Eq.~(\ref{Hamiltonian})].  The steady-state mean phonon number ($\bar{n}(\infty)$) is found with the Integration-free calculation explained in Appendix C. In Fig.~\ref{final phonon numbers of MRs wrt phase}, we plot the final mean phonon numbers of each of the mechanical resonators as a function of the closed-loop phase, $\phi_{\ell}$ for either of the exceptional-point coupling values, $|\mu|=\left|\mu_\mathrm{EP,1,2}\right|$. Either mechanical resonator can be cooled down to ground state selectively depending on $\phi_{\ell}$, which is indicative of the \textit{clockwise} versus \textit{counterclockwise} association of the $\phi_{\ell}=\pi/2$ and $3\pi/2$ values. That is, for $\phi_{\ell}=\pi/2$ and $3\pi/2$, the second or first mechanical resonator is respectively favored in cooling \cite{lai2020nonreciprocal,jiang2021energy}. In the same vein, when $\phi_{\ell}=0$ or $\pi$, resonator discrimination is lost as the time-reversal symmetry is restored. 
\begin{figure}[H]
  \centering
  \includegraphics[width=0.5\textwidth]{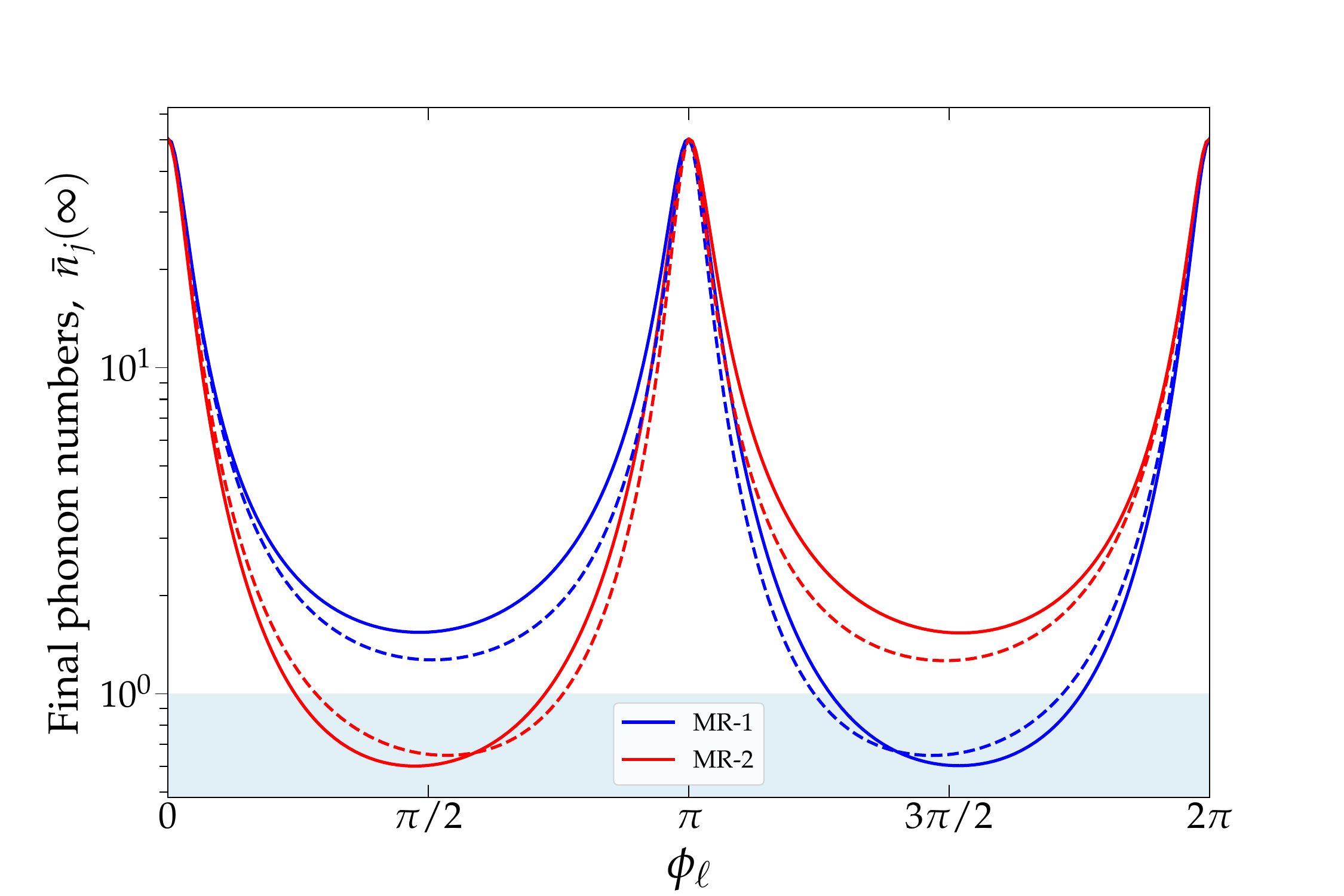}
\caption{Steady-state mean phonon numbers of the first (blue) and second (red) mechanical resonators as a function of $\phi_\ell$ computed at two EPs of $|\mu|=\left|\mu_\mathrm{EP,1}\right|$ (solid) and $|\mu|=\left|\mu_\mathrm{EP,2}\right|$ (dashed).}
\label{final phonon numbers of MRs wrt phase}
\end{figure}

\begin{figure}[H]
  \centering
  \includegraphics[width=0.5\textwidth]{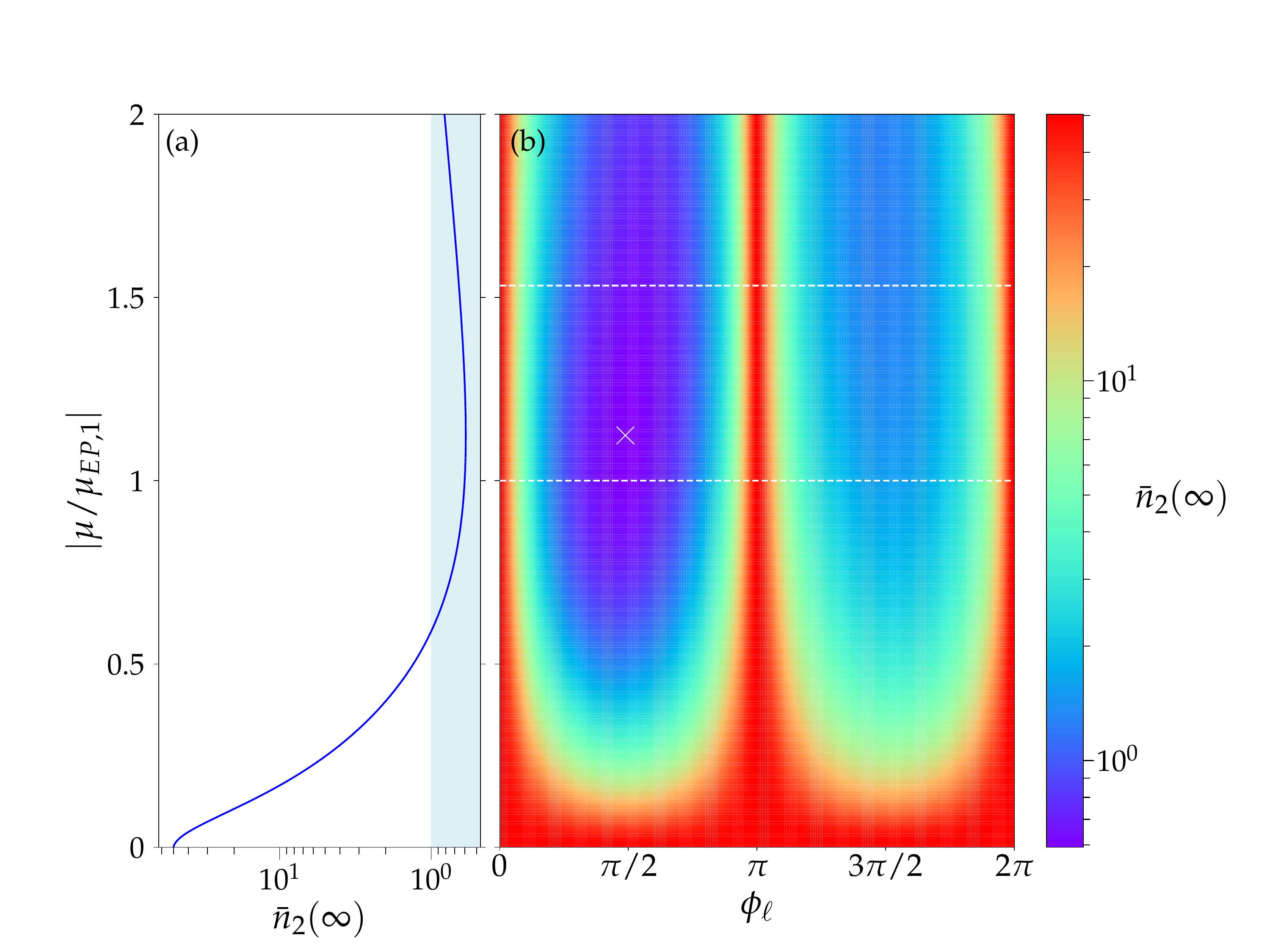}
\caption{Steady-state mean phonon number of the second mechanical resonator with respect to (a) $|\mu|$ where loop-phase is taken as $\phi_{\ell} \simeq 0.49 \pi$ which gives the minimum steady-state phonon number, marked with a cross on the right subplot and (b) with respect to $|\mu|$ and $\phi_{\ell}$. The horizontal dashed lines mark the first and second EPs, $\left|\mu_\mathrm{EP,1}\right|$ and $\left|\mu_\mathrm{EP,2}\right|$.}
\label{meanphononnumber-2d}
\end{figure}

Previous studies considered specific synthetic phases of $\phi_{\ell}=\pi/2$ and $3\pi/2$, and determined peak performance \textit{at} the EPs \cite{lai2020nonreciprocal,jiang2021energy}. This fact signals that non-Hermiticity in the system, which is responsible for the EPs, promotes cooling as well. One of these works explained this by relating EPs to field localization in the resonator with less loss \cite{liao2023exceptional,pengscience2014}, while the other study emphasized the unidirectionality of phonon transport for this scheme \cite{jiang2021energy}. 
To elaborate on these findings, we examine where the optimal mechanical cooling lies with respect to \textit{both} intermechanical resonator coupling and global phase. In Fig.~\ref{meanphononnumber-2d}(a), we plot mean phonon number of the second mechanical resonator at the $\phi_{\ell}\simeq 0.49\pi$ with respect to intermechanical coupling, which harbors the minimum phonon number as extracted from Fig.~\ref{meanphononnumber-2d}(b) which is marked with a cross. We observe that the minimum final mean phonon number of the second mechanical resonator [$\bar{n}_{2,min}(\infty) \simeq 0.591$] appears at $|\mu|=1.12|\mu_\mathrm{EP,1}|$ and $\phi_{\ell} \simeq 0.49\pi$, which is shifted from $\phi_{\ell}=\pi/2$, thus not exactly at an EP, unlike \cite{jiang2021energy} that registers optimal cooling \textit{at} the EP. 
As a side remark, much higher EP-induced cooling was reported for the \textit{open-loop} geometry \cite{liao2023exceptional}; however, it lacks mechanical resonator selectivity with the plaquette phase, as in here or \cite{jiang2021energy}.

\begin{figure}[H]
  \centering
  \includegraphics[width=0.5\textwidth]{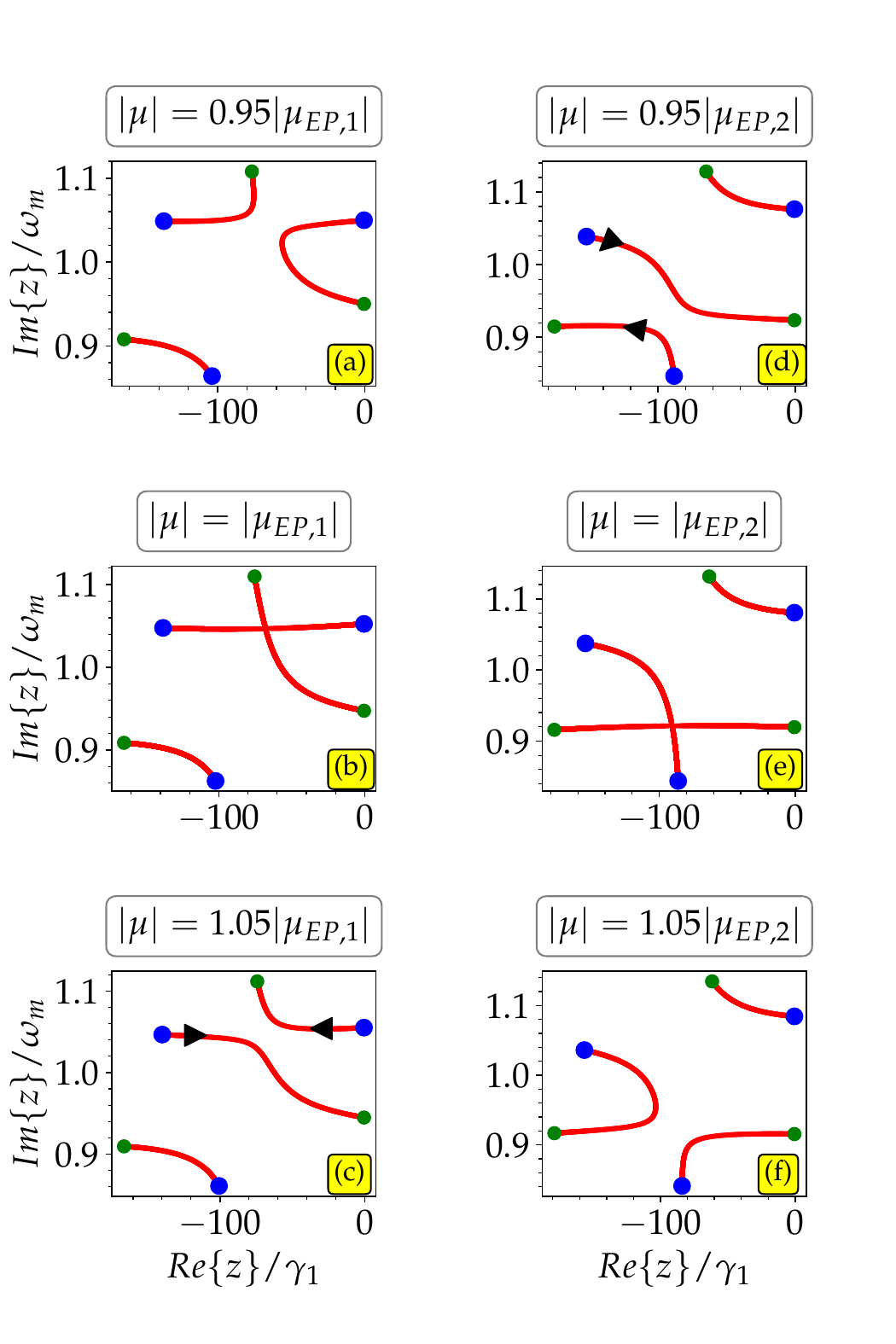}
\caption{Complex upper half-plane eigenvalue $z(|\mu|,\phi_{\ell})$ loci of the $\mathbf{M}^{(0)}$ matrix as $\phi_{\ell}$ is continuously varied between 0 (blue dots) to $\pi$ (green dots) at six different $|\mu|$ values: (a) 0.95 $\left|\mu_\mathrm{EP,1}\right|$,
(b) $\left|\mu_\mathrm{EP,1}\right|$, (c) 1.05 $\left|\mu_\mathrm{EP,1}\right|$, (d) 0.95 $\left|\mu_\mathrm{EP,2}\right|$, (e) $\left|\mu_\mathrm{EP,2}\right|$, (f) 1.05 $\left|\mu_\mathrm{EP,2}\right|$ .}
\label{root-loci}
\end{figure}

In order to gain more insight on why the range $\left|\mu_\mathrm{EP,1}\right|\le|\mu| \le\left|\mu_\mathrm{EP,2}\right|$ hosts optimal cooling, we plot in Fig.~\ref{root-loci} the complex upper half-plane eigenvalues $(z\left(|\mu|,\phi_{\ell}\right)=\alpha+i\omega,~\omega>0)$ of the stationary drift matrix, $\mathbf{M}^{(0)}$ at six different $|\mu|$ values while continuously varying $\phi_{\ell}\in [0,\pi]$; for $\phi_{\ell}\in [\pi,2\pi]$ the same loci are retraced backwards. The vertical axis designates oscillation angular frequency, and the horizontal axis the damping rate for $\alpha<0$. Subplots (b) and (e) in Fig.~\ref{root-loci} display the two EPs $\left|\mu_\mathrm{EP,1}\right|$ and $\left|\mu_\mathrm{EP,2}\right|$, where both real and imaginary parts of two modes as well as their eigenvectors coincide. Markedly, these occur at phase angles \textit{away} from $\pi/2$, unlike in our optomechanically induced transparency study \cite{beyza21}; unfortunately, this hampers an analytical treatment of the EPs. On the big picture, as $|\mu|$ is increased from below $\left|\mu_\mathrm{EP,1}\right|$ to above $\left|\mu_\mathrm{EP,2}\right|$ the bare modes undergo successive hybridizations, as we delineate below. Starting from $|\mu|\ll\left|\mu_\mathrm{EP,1}\right|$, as the intermechanical coupling $|\mu|$ is ramped up, the two originally degenerate phonon modes at $\omega_m$ split, forming lower- and upper-frequency phonon hybridizations. Concomitantly, the radiation pressure couplings 
between the optical and two mechanical resonators induce phonon-polariton characters.
For $|\mu|\sim\left|\mu_\mathrm{EP,1}\right|$, the upper-frequency polariton branch experiences an EP around 
1.04~$\omega_m$. As $|\mu|$ is increased to about $\left|\mu_\mathrm{EP,2}\right|$, this time the lower-frequency polariton 
branch EP is formed around 0.94~$\omega_m$. At either of these EPs, the two approaching skewed phonon-polariton states become completely parallel, which is the hallmark of non-Hermitian systems \cite{ozdemir2019parity}.
The range $|\mu| >\left|\mu_\mathrm{EP,2}\right|$ causes strongly intercoupled mechanical modes which are weakly coupled to the optical mode, opposite to the case in $|\mu| <\left|\mu_\mathrm{EP,1}\right|$. Thereby, the two extreme subplots (a) and (f) are the mirror reflections of one another with respect to a midway vertical line. Overall, the most critical aspect of the eigenvalue loci in Fig.~\ref{root-loci} for mechanical cooling purposes is the polariton-phonon mode coupling displayed with black arrows in plots (c) and (d), which links to the cold optical cavity mode. Once again, for $\phi_{\ell}<\pi/2$ ($\phi_{\ell}>\pi/2$), the second (first) mechanical resonator is preferentially cooled.

\begin{figure}[H]
  \centering
  \includegraphics[width=0.5\textwidth]{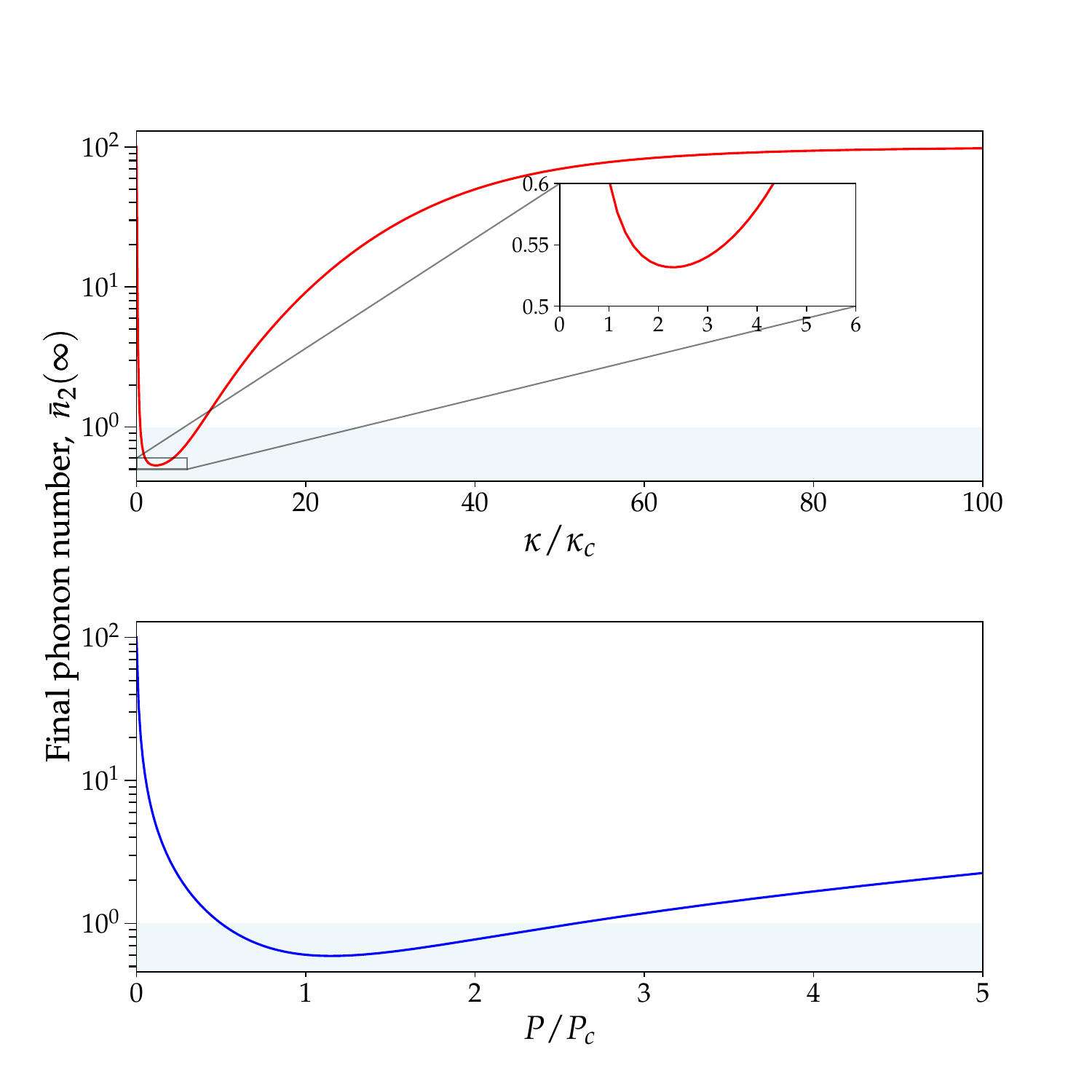}
\caption{Steady-state mean phonon number of the second mechanical resonator $(\phi_{\ell}=\pi/2)$ with respect to $\kappa$ and pump power, normalized to $\kappa_c/(2\pi)=$900~MHz, and $P_c = 0.25$~mW.}
\label{final phonon numbers wrt kappa and pump}
\end{figure}
Proceeding with the impact of the other parameters on the final mean phonon number,
in Fig.~\ref{final phonon numbers wrt kappa and pump}, we investigate cavity decay rate and pump power normalized to their values in the original parameter set $\kappa_c$ and $P_c$, respectively. The rapid increase in the final mean phonon number for $\kappa<\kappa_c$ (upper panel) arises from the fact that the photonic cavity becomes increasingly off-resonant compared to the broadening of the mechanical resonators' linewidths $\gamma_{1,2}$. On the other hand, the slow rise as $\kappa>\kappa_c$ is the manifestation of gradual departure from the beneficial resolved-sideband regime. Likewise, in the lower panel, we observe that pump power has an optimum value slightly above $P_c$. This sensitivity is due to the EP dependency on pump laser power. As the latter changes, the proximity to either of the EPs is adversely affected.

\begin{figure}[H]
  \centering
  \includegraphics[width=0.5\textwidth]{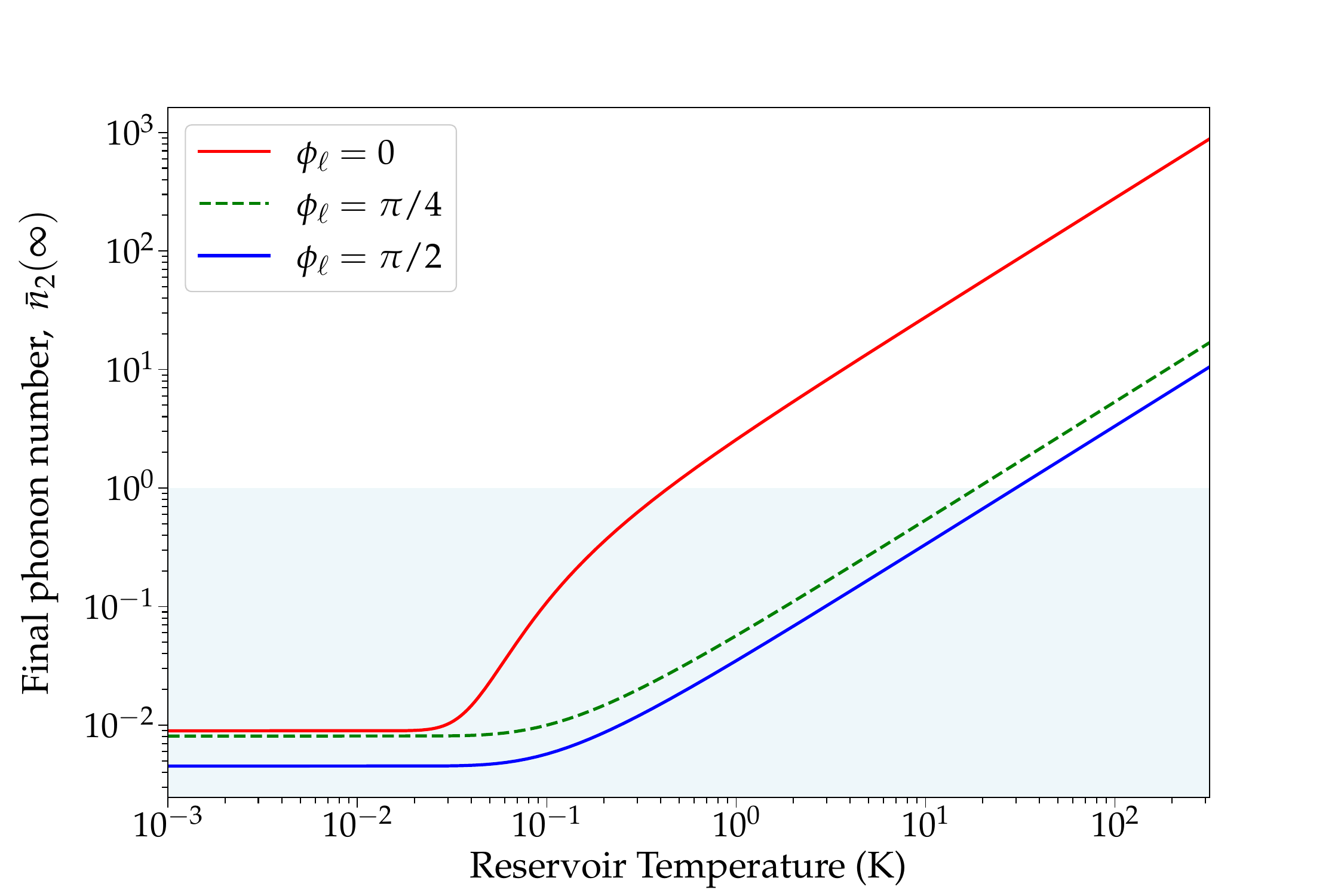}
\caption{Steady-state mean phonon number of the second mechanical resonator with respect to reservoir temperature at the exceptional point $|\mu_\mathrm{EP,1}|\simeq 52.5\,(\gamma_1+\gamma_2)$ for three different $\phi_\ell$ values.}
\label{final mean phonon numbers wrt temperature}
\end{figure}

In Fig.~\ref{final mean phonon numbers wrt temperature}, we plot the steady-state mean phonon number of the second mechanical resonator with respect to the ambient temperature (taken as equal for the cavity and mechanical resonators) from cryogenic to room temperature. In accordance with the Bose-Einstein distribution, the corner thermal energy beyond which $\bar{n}_2(\infty)$ rapidly increases corresponds to mechanical resonance, $\omega_m$. The ground-state cooling regime is marked by the shaded area. This is approached at higher temperatures depending on the closed-loop phase, with the optimal value being close to $\pi/2$ (see Fig.~\ref{final phonon numbers of MRs wrt phase}).

\subsection{Exceptional Surfaces}
To complement our parameter analysis, we bring in a recently perceived concept of an exceptional surface \cite{zhongprl19}. 
As we discussed in Figs.~\ref{meanphononnumber-2d} and \ref{root-loci}, having an EP has important physical consequences. However, 
if its presence is limited to a very finely chosen parameter set, it remains only a mathematical serendipity. In this regard, the formation of exceptional surfaces is highly welcomed to alleviate such experimental concerns \cite{zhongprl19}. 
\begin{figure}[H]
  \centering
   \hspace*{-0.7cm}
  \includegraphics[width=0.58\textwidth]{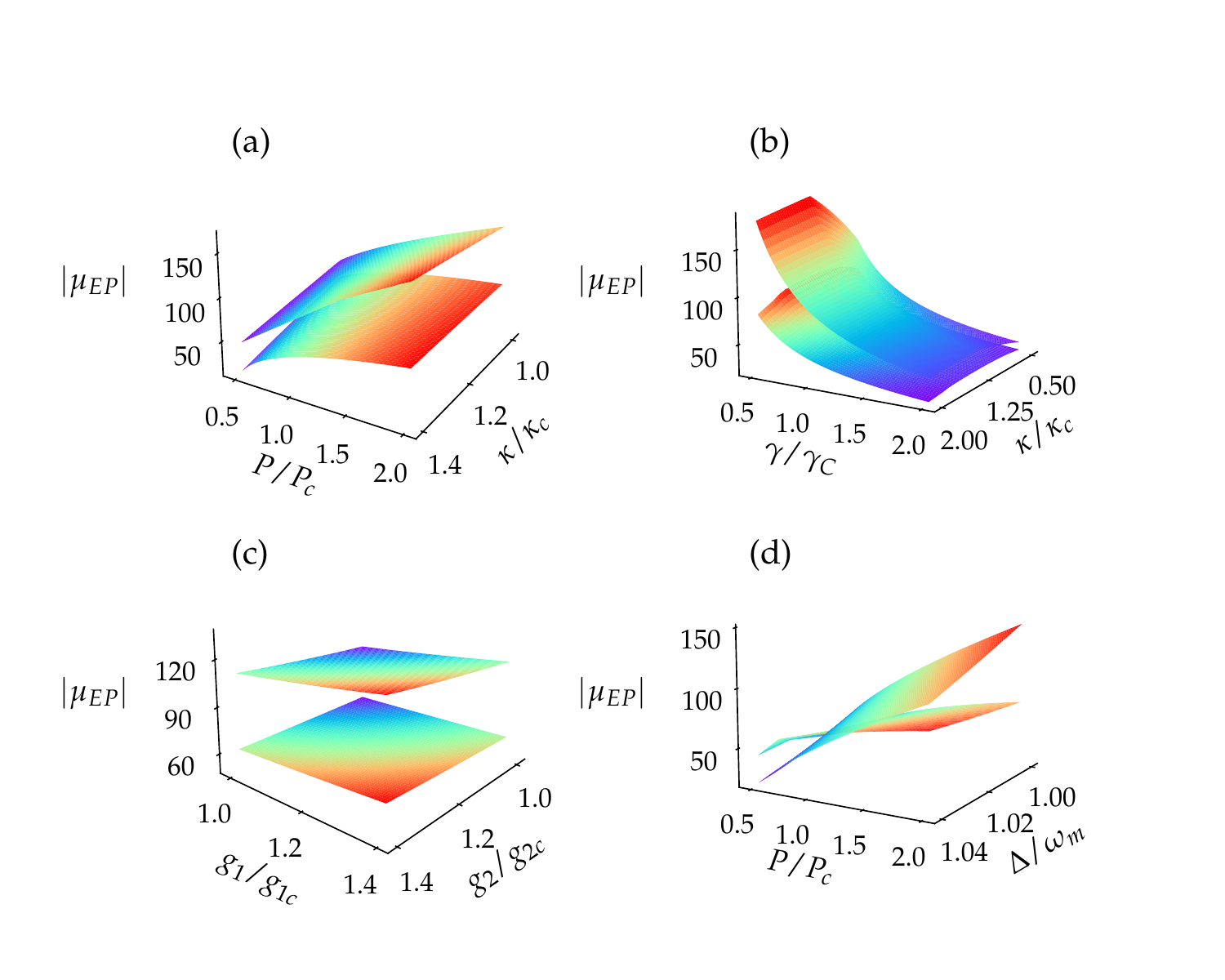}
\caption{Exceptional surfaces formed by first and second EPs where the vertical axis is ${|\mu_\mathrm{EP}|/(\gamma_1+\gamma_2)}$ with respect to a) $\kappa$ and pump power, b) $\kappa$ and $\gamma=\gamma_1=\gamma_2$, c) $g_1$ and $g_2$, and d) $\Delta$, and pump power. $\kappa_c$, $P_c$, $\gamma_c$, $g_{1c}$ and $g_{2c}$ corresponds to specific values of the cavity decay rate, pump power, mechanical losses ($\gamma_1 = \gamma_2$) and coupling constants stated in the parameter set. }
\label{Exceptional surfaces}
\end{figure}
Here, we systematically trace EPs over a range of different parameters to ratify that our system actually possesses exceptional surfaces. As depicted in Fig.~\ref{Exceptional surfaces}, we track the two EPs in relation to distinct parameter combinations. For the subplots (a)-(c), we see that they form a hypersurface contingent on the two specified parameters. As such, the system exhibits robustness against the variations of pump power, radiation pressure couplings, and mechanical losses so that it will reside on the exceptional surface. The exception to this is the behavior under detuning ($\Delta$) displayed in Fig.~\ref{Exceptional surfaces}~(d). The EPs are found only in a very restricted $\Delta$ range around $\omega_m$. In other words, a change in $\Delta$ corresponds to a dimension \textit{perpendicular} to the exceptional surface. Being limited to three-axis visualization, in Fig.~\ref{root-loci}, we can only plot the magnitudes of $\mu_\mathrm{EP}$. Because of this, in subplot (d), the two surfaces cross each other, which seemingly suggests the presence of a \textit{third-order} EP where all three eigenvalues coalesce \cite{hodaei2017enhanced}. However, upon further examination, we observed that in Fig.~\ref{Exceptional surfaces}~(d), two EPs always have different phases when their magnitudes are the same so that they each remain as second-order EPs. 
\subsection{Mechanical Squeezing}
\begin{figure}[H]
  \centering
  \includegraphics[width=0.5\textwidth]{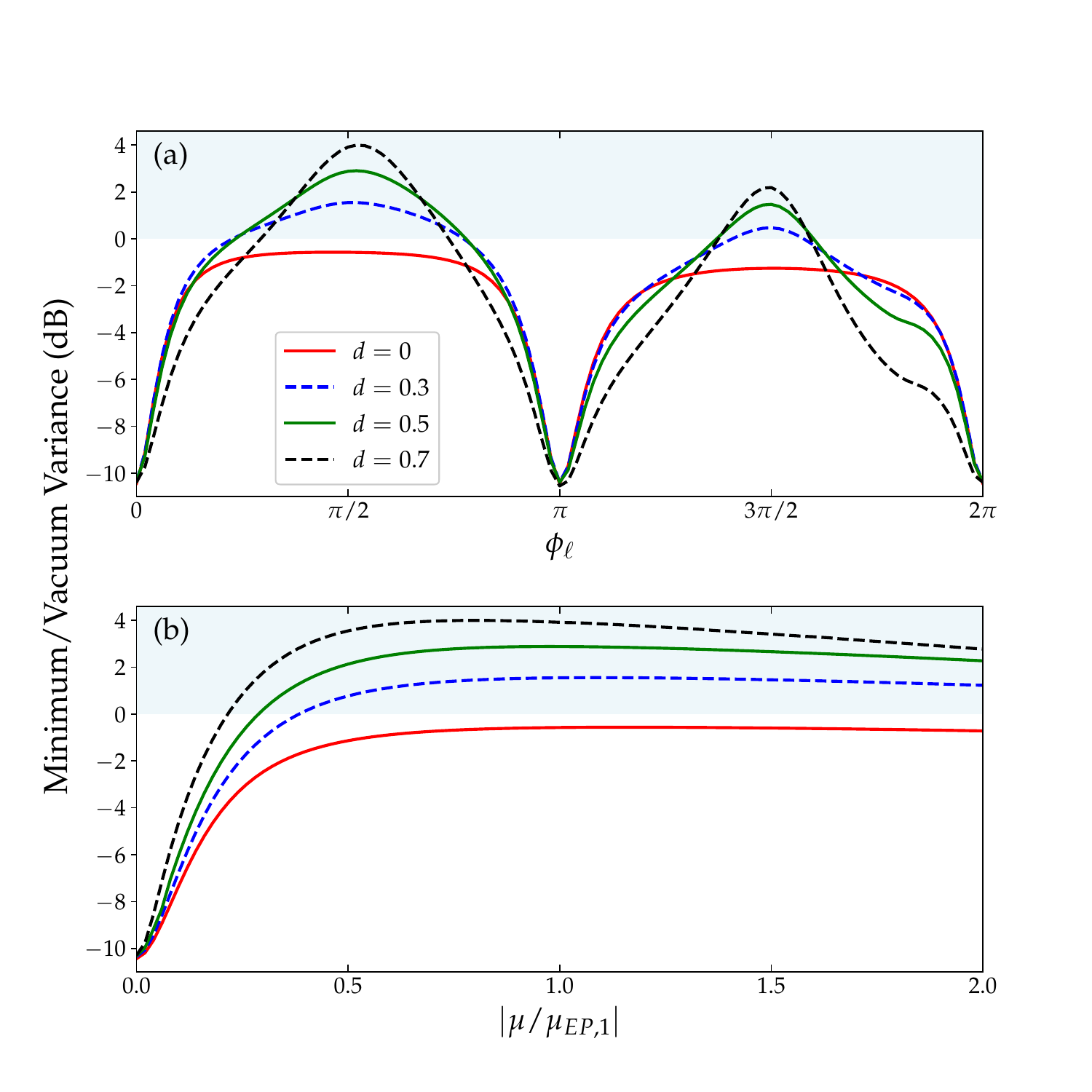}
\caption{Minimum variance with respect to vacuum value (in dB) of the second mechanical resonator (a) as a function of $\phi_{\ell}$ at $|\mu|\simeq 50.83\,(\gamma_1+\gamma_2)$, (b) as a function of mechanical coupling constant normalized to first exceptional point $\mu/\mu_{EP,1}$ at $\phi_{\ell}=\pi/2$ for different modulation depth values, $d=0, 0.3, 0.5, 0.7$; reservoir temperatures are taken as $T = 1.9$~K.}
\label{min var modulation}
\end{figure}

\begin{figure}[H]
  \centering
  \includegraphics[width=0.5\textwidth]{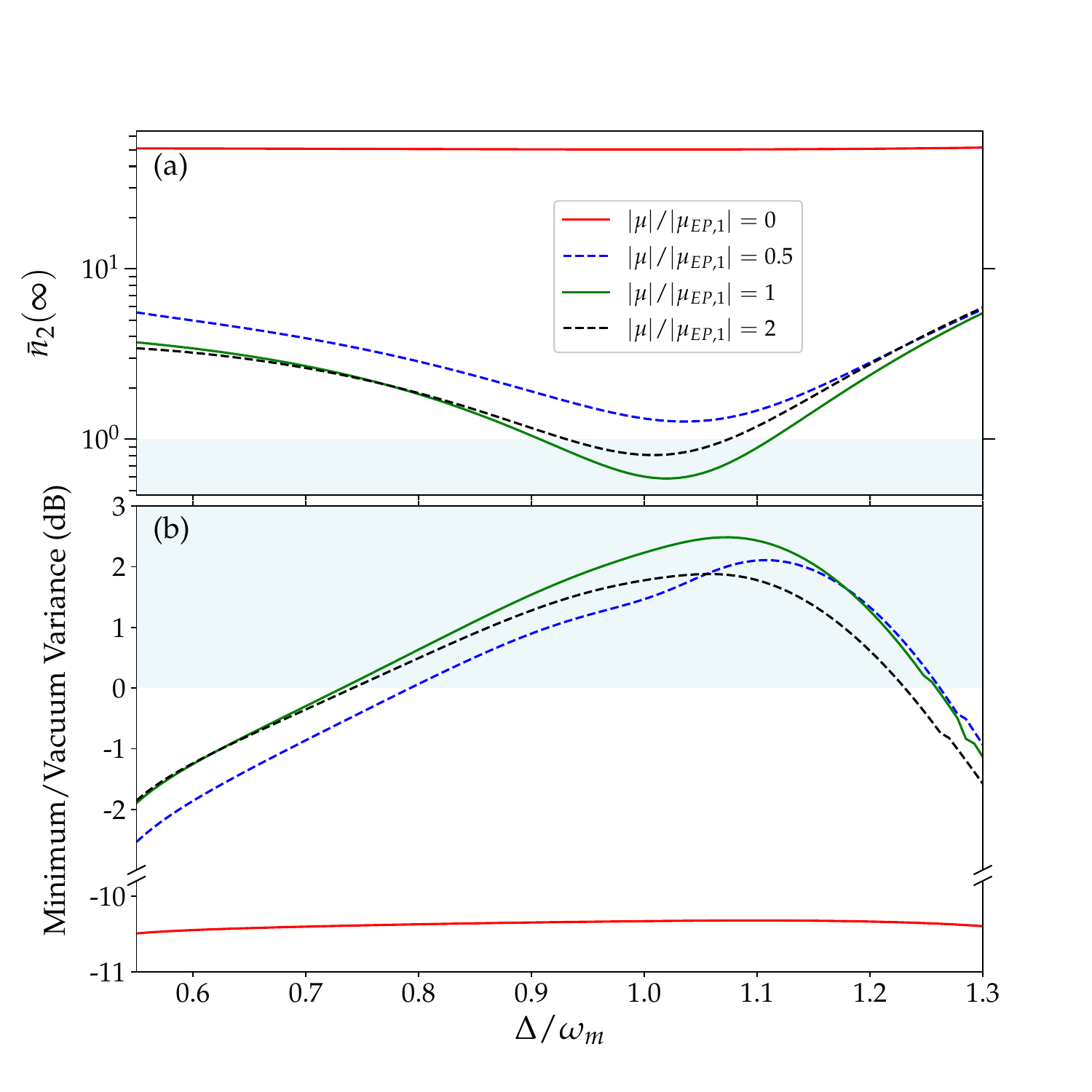}
\caption{(a) Steady-state mean phonon number of the second mechanical resonator at $|\mu_\mathrm{EP,1}|\simeq  52.5\,(\gamma_1+\gamma_2)$ in absence of modulation ($d = 0$), (b) minimum variance with respect to the vacuum value (in dB) of the second mechanical resonator at $|\mu_\mathrm{EP,1}|\simeq  51.36\,(\gamma_1+\gamma_2)$  with modulation depth $d = 0.4$ as a function of $\Delta$ normalized to mechanical frequency for different mechanical coupling constants; reservoir temperatures are taken as $T = 18.1$~K and $T = 1.9$~K, respectively.}
\label{frequencydependentsqueezing}
\end{figure}

To obtain mechanical squeezing beyond the vacuum level, we apply amplitude modulation over the pump laser \cite{mari2009gently}. We opt for the so-called \textit{upper} sideband modulation, i.e., $\epsilon_1\ne 0$, $\epsilon_{-1}= 0$. As with the laser carrier component, $\epsilon_0$ causing a shift in the EPs, a similar effect occurs under modulation. In particular, for the modulation depth of $d=\varepsilon_1/\varepsilon_0=0.5$, the first EP moves to $|\mu|\simeq 50.83\,(\gamma_1+\gamma_2)$.Mechanical squeezing is achieved when the variance is reduced below the vacuum value of 1/2, as marked by the shaded region in Fig.~\ref{min var modulation}. Here, we explore the effect of modulation depth, which is defined as the ratio $d=\varepsilon_1/\varepsilon_0$. The same control over the loop phase, $\phi_{\ell}$, also persists for the mechanical squeezing, displayed in Fig.~\ref{min var modulation}(a), and as expected, higher modulation depth enhances squeezing. Specifically, mechanical squeezing is reached for $d > 0.3$ (corresponding to 0~dB, marked with a dashed line), and towards $d \rightarrow 1$ system starts to become numerically unstable. As in cooling, the loop phase angle for maximum squeezing again occurs away from the $\pi/2$ value and is modulation depth-dependent. Fig.~\ref{min var modulation}(b) indicates that for $\mu\rightarrow 0$ the mechanical squeezing cannot be attained. As the mechanical coupling constant increases, so does the degree of squeezing. However, squeezing gradually diminishes as the system moves away from the first exceptional point.
 In Fig.~\ref{frequencydependentsqueezing}(a) and (b), we plot steady-state mean phonon number and the minimum variance of the second mechanical resonator as a function of detuning for different intermechanical coupling constants. Here, the parameter that breaks away from the exceptional surface, i.e., laser detuning, brings quite a beneficial feature, namely frequency-dependent cooling, and squeezing. In that respect, the importance of loop coupling is manifested by the absence of frequency dependence when intermechanical coupling is set to zero ($\mu=0$). In contrast, under both loop coupling and modulation, frequency-dependent squeezing is attained, which has recently gained importance, distinctly in the context of gravitational wave detection \cite{mcculler2020frequency}. As expected, maximum cooling and squeezing occurs close to $\mu_\mathrm{EP}$ and for the red-detuning of $\Delta\simeq \omega_m$ and $\Delta\simeq 1.1\omega_m$, respectively.
On either side, both the mechanical cooling and squeezing fall off as a function of the laser pump frequency. Beyond a certain detuning, cooling below ground state and squeezing below the vacuum level (marked by a shaded region) are lost.

\section{The validity range of the numerical techniques}
We would like to end our discussion on which extremes our numerical approaches can fail. Foremost, the computation of the steady-state variances hinges on the stability of the optomechanical response. For cooling, this is assured by having the eigenvalues of $\mathbf{M}^{(0)}$ on the left-hand complex plane \cite{beyza21}, as we displayed in Fig.~\ref{root-loci}. In the case of squeezing when the modulation depth, $d$, approaches one, we observed that the numerical convergence of the Floquet method becomes quite demanding even before the system crosses the instability. Seemingly, the integration-free computation also poses a potential limitation. Namely, as explained in Appendix~C, it relies on the diagonalization of $\mathbf{M}^{(0)}$, which is granted whenever this matrix is not defective. Therefore, exactly at EPs, it is not diagonalizable, with the closest alternative being the Jordan normal form \cite{lau2018fundamental}. Nevertheless, in practice, one can get arbitrarily near to EP while the technique works flawlessly. As a matter of fact, this is the case in Fig.~\ref{meanphononnumber-2d} where the horizontal dashed lines mark the first and second EPs, $\left|\mu_\mathrm{EP,1}\right|$ and $\left|\mu_\mathrm{EP,2}\right|$. Finally, the strong optomechanical-coupling regime though it is experimentally challenging \cite{verhagen-nature-2012} should not pose a numerical difficulty as neither of these techniques is perturbative in nature.

\section{Conclusion}
This study involves a theoretical exploration of closed-loop phase control within an elementary optomechanical plaquette operating in the quantum regime. It highlights the cooling and squeezing of the mechanical resonator selected according to the loop phase. We provide a physical insight into the crucial role of non-Hermiticity in proximity to the EPs, linking it to the behavior of complex eigenvalue loci influenced by the intermechanical resonator coupling. Another advantageous aspect is the presence of a parameter space containing an exceptional surface, ensuring the robustness of EP singularities under parameter variations. The laser detuning corresponds to the parameter causing deviation from the exceptional surface and facilitates frequency-dependent squeezing, a characteristic of practical significance. There are a number of directions in which this research can be further expanded. Foremost, the triangle plaquette can be enriched in the number of resonators and network topology, like being extended to a two-dimensional lattice with various synthetic flux patterns \cite{walternjp16}. Alternatively, the underlying $U(1)$ Abelian symmetry can be promoted to higher continuous symmetry groups or non-Abelian gauge potentials \cite{dalibardrmp11}. As opposed to these fundamental aspects, on the practical side, a timely pursuit is to assess the potential of the elementary optomechanical plaquette for quantum sensing and information processing purposes. Both of these require an in-depth analysis of quantum noise, specifically the role of thermal fluctuations, imprecision noise (output shot noise) and quantum back-action noise (radiation-pressure force noise) \cite{RevModPhys.82.1155}. These can be investigated in both cooling and squeezing settings to comprehend the quantum advantages in real-world sensing applications.

\begin{acknowledgments}
We are grateful to M. Paternostro, R. El-Ganainy, and C. Y\"uce for illuminating discussions.
\end{acknowledgments}
\begin{widetext}
\section*{Appendix~A: Steady-state values in quantum Langevin equations}
\setcounter{section}{1}
\renewcommand{\theequation}{A\arabic{equation}}
\setcounter{equation}{0}
\label{appendix:a}
The mean values of the cavity and mechanical modes satisfy
\begin{eqnarray*}
\label{mean a}
\frac{d\langle \hat{a} \rangle}{dt} &=& -i \Delta \langle \hat{a}(t) \rangle + \sqrt{\eta \kappa} \varepsilon_L(t) + i  \langle \hat{a}(t) \rangle(g_1\langle \hat{b}_1(t) \rangle^* + g_1^*\langle \hat{b}_1(t) \rangle) \\
\nonumber
& & +i \langle \hat{a}(t) \rangle (g_2 \langle \hat{b}_2(t)\rangle^* + g_2^* \langle \hat{b}_2(t) \rangle )-\frac{\kappa}{2} \langle \hat{a}(t) \rangle, \\
\label{mean b1}
\frac{d\langle \hat{b}_1 \rangle}{dt} &=& -i \omega_m \langle \hat{b}_1(t) \rangle + i\mu \langle \hat{b}_2(t)\rangle  + i g_1 |\langle \hat{a}(t) \rangle |^2 - \frac{\gamma_1}{2} \langle \hat{b}_1(t) \rangle, \\
\label{mean b2}
\frac{d\langle \hat{b}_2 \rangle}{dt} &=& -i \omega_m \langle \hat{b}_2(t) \rangle + i\mu^* \langle \hat{b}_1(t)\rangle  + i g_2 |\langle \hat{a}(t) \rangle |^2 - \frac{\gamma_2}{2} \langle \hat{b}_2(t) \rangle.
\end{eqnarray*}
In the steady-state, $\langle \hat{a}(t) \rangle $, $\langle \hat{b}_1(t) \rangle$ and $\langle \hat{b}_2(t) \rangle$ follow the input modulation period acting on the photonic cavity, $\varepsilon_L(t+\tau) = \varepsilon_L(t) = {\sum_{n=-\infty}^{\infty}\varepsilon_n}e^{-in\Omega t}$, \cite{teschl2012ordinary}. Here, we retain only the first neighboring sidebands, i.e., $e^{\pm i \Omega t}$. Expanding mean values into their harmonic components in the steady-state regime as $\langle \hat{\aleph} \rangle = \aleph_o + \aleph_1 e^{-i\Omega t} + \aleph_{-1} e^{i\Omega t},~\hat{\aleph}=\hat{a}, \hat{b}_1,\hat{b}_2$, coupled equations to be solved self-consistently for the center and sideband amplitudes are obtained as,
\begin{subequations}
\begin{eqnarray}
a_0 &=& \frac{\sqrt{\eta \kappa}\varepsilon_0-i(\Delta_{a,1}a_{-1} + \Delta_{a,-1}a_{1})}{i\Delta_{a,0}+\kappa/2},\\
a_{\pm 1} &=& \frac{\sqrt{\eta \kappa}\varepsilon_{\pm 1}-i\Delta_{a,\pm1}a_0}{i\Delta_{a,0}+\kappa/2 \mp i\Omega},\\
b_{1,0} &=& \frac{i\mu b_{2,0}+i g_1(|a_0|^2+|a_1|^2+|a_{-1}|^2)}{i\omega_m+\gamma_1/2},\\
b_{1,\pm 1} &=& \frac{i\mu b_{2,\pm 1}+i g_1(a_0 a_{\mp 1}^*+ a_0^* a_{\pm 1})}{i\omega_m\mp i\Omega+\gamma_1/2},\\
b_{2,0} &=& \frac{i\mu^* b_{1,0}+i g_2(|a_0|^2+|a_1|^2+|a_{-1}|^2)}{i\omega_m+\gamma_2/2},\\
b_{2,\pm 1} &=& \frac{i\mu^* b_{1,\pm 1}+i g_2(a_0 a_{\mp 1}^*+ a_0^* a_{\pm 1})}{i\omega_m\mp i\Omega+\gamma_2/2},
\end{eqnarray}
\label{sideband-amplitudes}
\end{subequations}
where $\Delta_a = \Delta-2\operatorname{Re}\left[g_1 \langle \hat{b}_1(t)\rangle^* + g_2 \langle \hat{b}_2(t)\rangle^*\right]$ is the detuning which is indirectly modulated with the motion of mechanical modes. Likewise, this detuning can be separated into its harmonics as $\Delta_a (t) = \Delta_{a,0}+ \Delta_{a,-1} e^{i \Omega t}+\Delta_{a,1} e^{-i \Omega t}$, where the coefficients of harmonics are found as
\begin{eqnarray*}
\Delta_{a,0} & = &  \Delta-2\operatorname{Re}(g_1 b_{1,0}^* + g_2 b_{2,0}^*),\\
\Delta_{a,1} & = &  -(g_1^* b_{1,1} + g_1 b_{1,-1}^* + g_2 b_{2,1} + g_2 b_{2,-1}^* ),\\
\Delta_{a,-1} &=& \Delta_{a,1}^*.
\end{eqnarray*}

\section*{Appendix~B: Covariance matrix within Floquet formalism}
\label{appendix:b}
\renewcommand{\theequation}{B\arabic{equation}}
\setcounter{equation}{0}
We extend this formalism to the covariance matrix $\mathbf{V}(t)=\langle \mathbf{\hat{R}}(t)  \mathbf{\hat{R}}^T(t) \rangle $ consisting of periodic entries in the steady state so that we can expand $V_{ij}(t)=\langle \hat{R}_i(t)  \hat{R}_j(t) \rangle$ in Fourier series as 
\begin{eqnarray*}
V_{ij}(t) &=& \sum_{\ell} e^{-i\ell \Omega t} V_{ij}^{(\ell)}(t) = \sum_{m,m'}  e^{-i(m+m') \Omega t} \langle \hat{R}^{(m)}_i(t)  \hat{R}^{(m')}_j(t) \rangle, \\
& = & \sum_{\ell} e^{-i\ell \Omega t} \sum_{m}  \langle \hat{R}^{(m)}_i(t)  \hat{R}^{(\ell-m)}_j(t) \rangle, 
\end{eqnarray*}
where $ \hat{R}^{(m)}_i(t)=\frac{1}{2\pi} \int_{-\infty}^{\infty} \hat{R}^{(m)}_i(\omega) e^{-i\omega t} d\omega $. Setting $j=i$ since we are interested in the variance, and we obtain
\begin{equation}
\label{variancefourier}
\langle  \hat{R}^{(m)}_i(t)  \hat{R}^{(\ell-m)}_i(t) \rangle = \frac{1}{4\pi^2} \int_{-\infty}^{\infty} d\omega \int_{-\infty}^{\infty} d\omega' e^{-i(\omega+\omega')t} \langle \hat{R}^{(m)}_i(\omega)  \hat{R}^{(\ell-m)}_i(\omega') \rangle.
\end{equation}
For simplicity, we switch to global matrix indices by avoiding the Fourier component index i.e., $\hat{R}^{(m)}_i \rightarrow R_p ,~\hat{R}^{(\ell-m)}_i \rightarrow R_q$ where index $i$ stands for cavity and mechanical modes and index $m=[-N,N]$ denotes the number of zones retained in Floquet expansion. Then, after combining the Fourier index and quadrature mode indices, we have  
\begin{eqnarray}
\langle  \hat{R}^{(m)}_i(\omega)  \hat{R}^{(\ell-m)}_i(\omega) \rangle  = \langle  \hat{R}_p(\omega)  \hat{R}_q(\omega') \rangle = \sum_{p',q'} T_{pp'}(\omega) T_{qq'}(\omega')   \langle  \hat{n}_{p'}(\omega)  \hat{n}_{q'}(\omega')  \rangle,
\end{eqnarray}
where $\hat{R}_p = T_{pp'}\hat{n}_{p'}$ and $\hat{R}_q = T_{qq'}\hat{n}_{q'}$. Only non-zero contribution for noise correlation term comes from only stationary Fourier component, $\ell=0$. Previously, we defined $\mathbf{K}(\tau,\tau') =  \langle  \hat{n}_{i}(\tau)  \hat{n}_{j}(\tau')  \rangle = \mathbf{C} \delta(\tau-\tau')$. Then, the noise correlation term becomes 
\begin{eqnarray}
\langle  \hat{n}_{p'}(\omega)  \hat{n}_{q'}(\omega')  \rangle = \int_{-\infty}^{\infty} \int_{-\infty}^{\infty} dt dt' e^{i\omega t} e^{i\omega' t'} \langle  \hat{n}_{p'}(t)  \hat{n}_{q'}(t')  \rangle,    
\end{eqnarray}
with $\langle  \hat{n}_{p'}(t)  \hat{n}_{q'}(t')  \rangle = \delta(t-t') C_{p',q'}^{(0)} $.
In the end, we have 
\begin{eqnarray}
\langle  \hat{n}_{p'}(\omega)  \hat{n}_{q'}(\omega')  \rangle  = 2\pi C_{p',q'}^{(0)} \delta(\omega+\omega').
\end{eqnarray}
We no longer have time dependence on the right-hand side and invoking the steady state for reaching periodicity i.e., $t \rightarrow \infty$
\begin{eqnarray*}
\langle  \hat{R}_p(\infty)  \hat{R}_q(\infty) \rangle &=& \frac{1}{2\pi} \int_{-\infty}^{\infty} d\omega \sum_{p',q'} T_{pp'}(\omega) C_{p'q'}^{(0)} T_{q'q}(-\omega),    \\
& = &\frac{1}{2\pi}  \int_{-\infty}^{\infty} d\omega\, \mathbf{T}(\omega)~ \mathbf{C} ~\mathbf{T}^{T}(-\omega).
\end{eqnarray*}
We need to revert back from $p,q$ to $i,m$ indices, variance in steady state becomes
\begin{equation}
V_{ii}^{(\ell)}(\infty) = \sum_{m=-N}^{N} \langle  \hat{R}_i^{(m)}(\infty)  \hat{R}_i^{(\ell-m)}(\infty) \rangle.   
\end{equation}
We will further simplify the variance equation by using the definition of the covariance matrix.
\begin{equation}
V_{ii}^*(t)= \sum_{\ell=-\infty}^{\ell=\infty} e^{i\ell \Omega t} (V_{ii}^{(\ell)})^{*}=  \sum_{\ell=-\infty}^{\ell=\infty} e^{-i\ell \Omega t} (V_{ii}^{(\ell)}),    
\end{equation}
then we have $V_{ii}^{(\ell)} =(V_{ii}^{(-\ell)})^{*}$ since the diagonal components of covariance matrix are real. Using this equality for Fourier components, 
\begin{eqnarray*}
V_{ii}(t) &=& V_{ii}^{(0)} + \sum_{\ell=1}^{\ell=\infty} V_{ii}^{(\ell)}  e^{-i\ell \Omega t} + (V_{ii}^{(\ell)})^{*} e^{i\ell \Omega t},\\
& = & V_{ii}^{(0)} + \sum_{\ell=1}^{\ell=\infty} V_{ii}^{(\ell)} 2\operatorname{Re}(V_{ii}^{(\ell)} e^{-i\ell \Omega t}), \\
& = & V_{ii}^{(0)} + 2\sum_{\ell=1}^{\ell=\infty} |V_{ii}^{(\ell)}| \cos(\ell \Omega t -\phi_{ii}^{\ell}).
\end{eqnarray*}
We set $\ell_{max}=1$  since $|V_{ii}^{(2)}| \ll V_{ii}^{(1)}$, then maximum squeezing corresponding to minimum variance is 
\begin{equation}
\operatorname{min}(V_{ii}(t)) = V_{ii}^{(0)}- 2|V_{ii}^{(1)}|.   
\end{equation}

\section*{Appendix~C: Integration-free computation of steady-state variances}
\label{appendix:c}
\renewcommand{\theequation}{C\arabic{equation}}
\setcounter{equation}{0}
In this appendix, we consider the case where there is no modulation so that Eq.~(\ref{inf matrix}) reduces to 
\begin{equation}
\left[i\omega\mathbf{I} + \mathbf{M}^{(0)}\right] \mathbf{\hat{R}}(\omega)=-\mathbf{\hat{N}}(\omega)\, .
\end{equation}
As $\mathbf{M}^{(0)}$ has no $\omega$ dependence, for the integration-free computation of steady-state variances, our aim is to extract the explicit frequency dependence of $\mathbf{T}(\omega)$ when the above equation is inverted to 
\begin{equation}
\mathbf{\hat{R}}(\omega)=-\mathbf{T}(\omega)\,\mathbf{\hat{N}}(\omega)\, .
\end{equation}
This is achieved by first applying a similarity transformation that diagonalizes $\mathbf{M}^{(0)}$ as
\begin{equation}
\mathbf{U}^{-1}\mathbf{M}^{(0)}\mathbf{U}=\mathrm{diag}[\lambda_1,\cdots,\lambda_i,\cdots,\lambda_6],
\end{equation}
where $\mathbf{U}$ is simply composed of the eigenvectors of $\mathbf{M}^{(0)}$, with the corresponding eigenvalues 
being $\{\lambda_i\}$. After this diagonalization, matrix inverse is trivially performed, which is followed by undoing the similarity 
transformation, leading to the frequency dependence of the entries of $\mathbf{T}$ matrix as
\begin{equation}
T_{ij}(\omega)=\sum_{k}\frac{U_{ik}U^{-1}_{kj}}{i\omega+\lambda_k}\,.
\end{equation}
With this explicit form, we can perform the integration for the steady-state covariance matrix
\begin{equation}
\mathbf{V}(\infty)=\frac{1}{2\pi}  \int_{-\infty}^{\infty} d\omega\, \mathbf{T}(\omega)~ \mathbf{C} ~\mathbf{T}^{T}(-\omega)\, ,
\end{equation}
using the residue calculus, which yields the entries
\begin{equation}
V_{ij}(\infty)= \sum_{l,q}\sum_{k,k'} U_{ik} U^{-1}_{kl} C_{lq} U^{-1}_{k'q} U_{jk'} 
\frac{\mathrm{sgn(\mathrm{Re}\{\lambda_k\})}+\mathrm{sgn(\mathrm{Re}\{\lambda_k'\})}}{2\left(\lambda_k+\lambda_k'\right)} \, .
\end{equation}
Here, $\mathrm{sgn(\cdot)}$ is the sign function, and $\mathrm{Re}\{\cdot\}$ extracts the real part.

As an alternative expression, in the literature, the so-called \textit{symmetrically ordered} covariance matrix 
is widely used, which has the entries
\begin{equation}
V^s_{ij}(t)=\frac{1}{2}\langle\hat{R}_i(t)\hat{R}_j(t)+\hat{R}_j(t)\hat{R}_i(t)\rangle\, ,
\end{equation}
and it satisfies the following Ricatti/Lyapunov matrix differential equation \cite{mari2009gently,woolley2014two,pietikainen2020combining,hanpra2019,liao2023exceptional}
\begin{equation}
\label{lyapunov}
\mathbf{\dot{V^s}}= \mathbf{M}^{(0)} \mathbf{V^s} + \mathbf{V^s} \mathbf{M}^{(0)T} + \mathbf{D}\, ,
\end{equation}
where 
\begin{equation}
\mathbf{D}=\frac{1}{2}\mathrm{diag}[\kappa(2n_a+1),\kappa(2n_a+1),\gamma_1(2n_m+1),\gamma_1(2n_m+1),\gamma_2(2n_m+1),\gamma_2(2n_m+1)]\,.
\end{equation}
By setting to zero the left-hand side of Eq.~(\ref{lyapunov}) for the steady state, and performing the same similarity 
transformation on $\mathbf{M}^{(0)}$, we obtain an even simpler expression as
\begin{equation}
V^s_{ij}(\infty)= \sum_{k,q} U_{ik} \frac{-D^\prime_{kq}}{\lambda_k+\lambda_q} U_{jq} \, .
\end{equation}
where $\mathbf{D}^\prime=\mathbf{U}^{-1}\mathbf{D}\left(\mathbf{U}^{-1}\right)^T$.
\section*{Data Availability Statement}
No data are associated with this manuscript.
\end{widetext}

\end{document}